\documentclass[twocolumn,aps,prb,amsmath,amssymb]{revtex4-2}
\usepackage{bm}
\usepackage{amsmath}
\usepackage{amssymb}
\usepackage{graphicx}
\usepackage[unicode=true,
 bookmarks=false,
 breaklinks=false,pdfborder={0 0 1},backref=false,colorlinks=false]
 {hyperref}
\hypersetup{
 colorlinks}

\makeatletter

\usepackage[varg]{txfonts}
\usepackage{color}
\usepackage{tikz}
\usetikzlibrary{shapes}
\usepackage{xcolor}

\usepackage{bm}

\newcommand{\bigDiamond}{\mathop{\mathpalette\bigDi@mond\relax}}
\newcommand{\bigDi@mond}[2]{\vcenter{\hbox{\m@th \scalebox{\ifx#1\displaystyle 2\else1.2\fi}{$#1\Diamond$}}}}
\newcommand{\RNum}[1]{\uppercase\expandafter{\romannumeral #1\relax}}

\def\XXint#1#2#3{{\setbox0=\hbox{$#1{#2#3}{\int}$}
    \vcenter{\hbox{$#2#3$}}\kern-.5\wd0}}

\def\be{\begin{equation}}
\def\ee{\end{equation}}

\def\bi{\begin{itemize}}
    \def\ei{\end{itemize}}
\def\bn{\begin{enumerate}}
    \def\en{\end{enumerate}}
\def\bea{\begin{eqnarray}}
\def\eea{\end{eqnarray}}
\newcommand{\bpm}{\begin{pmatrix}}
    \newcommand{\epm}{\end{pmatrix}}

\def\ba{\begin{array}}
    \def\ea{\end{array}}
\def\bd{\begin{displaymath}}
\def\ed{\end{displaymath}}

\renewcommand{\imath}{\hspace{1pt}\mathrm{i}\hspace{1pt}}

\makeatother

\begin{document}
\title{Spinon Kondo lattice in quantum spin liquids }
\author{Xia-Ming Zheng}
\affiliation{Department of Physics, Sharif University of Technology, Tehran 14588-89694,
Iran}
\author{Mehdi Kargarian}
\email{kargarian@sharif.edu}

\affiliation{Department of Physics, Sharif University of Technology, Tehran 14588-89694,
Iran}
\begin{abstract}

Motivated by recent experimental observations of Kondo resonances
in cobalt atoms on single layer 1T-TaSe$_{2}$, we theoretically investigate
the effect of coupling a U(1) quantum spin liquid with a spinon Fermi
surface to a lattice of Anderson impurities. Within the slave-rotor
formalism, we find that above a critical coupling strength between
the spin liquid and impurity lattice, the spinons hybridize to form
heavy quasiparticles near the Fermi level, realizing a \textit{spinon
Kondo lattice phase} analogous to heavy fermion materials. Using the
Bethe-Salpeter equation and accounting for emergent gauge fluctuations,
we compute the spectral density and density of states, revealing the
formation of spinon-chargon bound states in the spinon Kondo lattice
phase. We characterize the thermodynamic and spectroscopic signatures
of this phase, demonstrating specific heat and neutron scattering
responses distinct from a pure quantum spin liquid. Our findings establish
the spinon Kondo lattice as a framework to study the rich physics
of spin liquids. 
\end{abstract}
\maketitle

\section{Introduction}


Quantum spin liquids (QSL) and resonating valence bond (RVB) states
were initially introduced by Anderson as a physical mechanism to explain
high $T_{c}$ superconductivity in cuprates \citep{Anderson1973,Anderson1987}.
QSL is a quantum paramagnetic Mott insulator phase which evades a
long-range magnetic order even at zero temperature. Instead, the QSL
is an exotic quantum phase of quantum matter described by topological
order \cite{Wen2002,Wen2004} and many-body long-range entanglement
\citep{Chen2010}. The emergent gauge fields and symmetry classification
by projective symmetry group are used to reveal the existence of a
variety of quantum phases with the same symmetry yet with distinct
properties. Among all, the U(1) quantum spin liquid with spinon Fermi
surface is a spin liquid where spinons are gapless forming a Fermi
surface and are coupled with emergent U(1) gauge fields \citep{Lee2005,Lee2006}.
Due to the presence of spin-charge separation in Mott insulators,
the elementary dynamical units are fractionalized to spinons and chargons
\cite{Lee1998,Florens2002,Florens2004,Lee2005}. The spinons are charge
neutral spin-1/2 fermion and chargons are spinless boson with electron
change $-e$. Therefore, this U(1) quantum spin liquid and some of
their physical characteristics resemble those of Fermi liquids. For
example, the entanglement entropy in real space follows a logarithmic
area law $S_{E}\sim L\log(L)$, where $L$ represents the boundary
length, for both the U(1) quantum spin liquid and Fermi liquid \citep{Swingle2010,Zhang2011}.
On the other hand, the coupling of conventional metals to localized
magnetic moments gives rise to the Kondo bound state and the Kondo
effect \citep{Anderson1961,Kondo1964,Hewson1997}, characterized by
a prominent Kondo resonance peak in the electronic spectrum.

The recent experimental observation of resonant states in cobalt atoms
on single-layer 1T-TaSe$_{2}$ provided evidence of the spinon Kondo
effect in a spin liquid \citep{Chen2022}. It has been found that
the coupling of the U(1) quantum spin liquid with impurities leads
to the emergence of the spinon Kondo effect \citep{He2022}. Although
the spin liquid is an insulator, it exhibits properties similar to
the electronic Kondo effect, with the spinon spectrum resembling the
electron spectrum in the Kondo effect. Resonance peaks appear at the
inner edges of the upper and lower Hubbard bands in the spin liquid,
which are attributed to the formation of spinon-chargon bound states
induced by the emergent gauge fields near the impurity. Therefore,
despite the complexity of the U(1) quantum spin liquid compared to
the Fermi liquid, it often exhibits similarities due to the presence
of a Fermi surface in its internal dynamical units.

The Kondo lattice model is formed by coupling a Fermi liquid to a
lattice of magnetic ions. The hybridization between the latter and
the host electrons dissolves the impurity spins into the Fermi liquid,
endowing them with electric charge \citep{Coleman2005}. This process
gives rise to the formation of heavy fermion quasiparticles \citep{Coleman1983,Hewson1997,Coleman2015},
known as heavy Fermion materials, characterized by a significant increase
in the effective mass of the quasiparticles. Furthermore, if the magnetic
interaction between ions is taken into account and assuming they form
a spin liquid, a phase transition from a heavy fermion phase with
a large Fermi surface to a fractionalized Fermi liquid (FL$^{*}$)
phase occurs as the Kondo interaction strength decreases \citep{Burdin2002,Senthil2004,Sachdev2023}.

Motivated by recent scanning tunneling spectroscopy measurements on
single-layer 1T-TaSe$_{2}$ and the observation of Kondo resonances
\citep{Chen2022}, we speculate that similar phenomena to the Kondo
lattice may arise when a quantum spin liquid is coupled with a lattice
of magnetic ions. The main message of our work is to introduce and
convey the concept of the $\textit{spinon Kondo lattice phase}$.
We introduce a model consisting of a quantum spin liquid coupled to
a lattice of Anderson impurities (AL) and explore its connection to
heavy fermion effects. In particular, we aim to answer the following
questions: i) how does the Anderson impurity lattice affect the spinon
Fermi surface and the single-particle spectra of spinons and chargons?
ii) does the coupling between the impurity lattice and the U(1) quantum
spin liquids give rise to new quantum phases? iii) what is the influence
of emergent gauge fields in the U(1) quantum spin liquids on the system?
iv) what are the possible experimental signatures of phases? To address
these questions, we organize the paper as follows. Sec. \ref{sec II}
begins with the Hubbard model coupled with Anderson impurity lattice
and analyzes the phases using the slave rotor method and mean-field
approximation, elucidating the quantum phase transitions and parton
single-particle spectra. In Sec. \ref{sec III}, we investigate the
physical electron excitations through parton Green's functions and
the Bethe-Salpeter equation, taking into account the effects of emergent
gauge fields on the Green's functions and single-electron spectrum.
In Sec. \ref{sec IV}, we explore the thermodynamic properties and
relevant observables of our model. The Sec. \ref{conclusions} summarizes
the main findings. The details of derivations of some expressions
are relegated to Appendices.

\section{Slave rotor approach to Anderson impurity lattice\label{sec II}}


The Hamiltonian of the Hubbard model, coupled to an impurity lattice
\citep{Hewson1997}, in the spin liquid phase can be expressed as
follows:

\begin{align}
H= & \sum_{i,j,\sigma}t_{ij}c_{i,\sigma}^{\dagger}c_{j,\sigma}+\sum_{i\sigma}\epsilon_{d}d_{i,\sigma}^{\dagger}d_{i,\sigma}+V\sum_{i\sigma}c_{i,\sigma}^{\dagger}d_{i,\sigma}+\mathrm{h.c.}\nonumber \\
 & +\frac{U_{\mathrm{QSL}}}{2}\sum_{i}(n_{ic}-1)^{2}+\frac{U}{2}\sum_{i}(n_{id}-1)^{2},\label{original H}
\end{align}

\noindent where $c(d)$ are fermionic annihilation operators of electrons
residing on sites of the lattice of itinerant electrons (Anderson
impurities), and the corresponding number operators are $n_{ic}=\sum_{\sigma}c_{i,\sigma}^{\dagger}c_{i,\sigma}$
and $n_{id}=\sum_{\sigma}d_{i,\sigma}^{\dagger}d_{i,\sigma}$. In
this expression, $t_{ij}$ is the hopping integral, $\epsilon_{d}$
is the energy of the impurity electron, $\sigma=\{\uparrow,\downarrow\}$
is the spin index, $V$ is the strength of the coupling between the
itinerant and impurity electrons. $U_{\mathrm{QSL}}$ is the Hubbard
interaction between host itinerant electrons, and we assume that it
is strong enough to drive the host system into a spin liquid phase.
$U$ is the Coulomb repulsion between electrons on a single Anderson
impurity.

We utilize the slave rotor construction \citep{Florens2002,Florens2004}
to express the electron operators as composites of spinon and chargon
operators: $c_{i,\sigma}=f_{i,\sigma}X_{i}^{\dagger}$ and $d_{i,\sigma}=a_{i,\sigma}Y_{i}^{\dagger}$,
where $X_{i}=e^{-i\theta_{i}}$ and $Y_{i}=e^{-i\phi_{i}}$ represent
the field operators of spinons and chargons, respectively. Substituting
these relations into Eq. \eqref{original H}, we obtain:

\begin{align}
H= & \sum_{i,j,\sigma}t_{ij}f_{i,\sigma}^{\dagger}f_{j,\sigma}X_{j}^{\dagger}X_{i}+\mathrm{h.c.}-\sum_{i\sigma}\left(\mu+h_{1,i}\right)f_{i,\sigma}^{\dagger}f_{i,\sigma}\nonumber \\
 & +\sum_{i\sigma}\left(\epsilon_{d}-h_{2,i}\right)a_{i,\sigma}^{\dagger}a_{i,\sigma}+V\sum_{i\sigma}f_{i,\sigma}^{\dagger}a_{i,\sigma}Y_{i}^{\dagger}X_{i}+h.c.\nonumber \\
 & +U_{\mathrm{QSL}}\sum_{i}P_{i}^{\dagger}P_{i}+i\sum_{i}h_{1,i}P_{i}X_{i}-i\sum_{i}h_{1,i}X_{i}^{\dagger}P_{i}^{\dagger}\nonumber \\
 & +U\sum_{i}Q_{i}^{\dagger}Q_{i}+i\sum_{i}h_{2,i}Q_{i}Y_{i}-i\sum_{i}h_{2,i}Y_{i}^{\dagger}Q_{i}^{\dagger}\nonumber \\
 & +\sum_{i}\lambda_{1,i}\left(X_{i}^{\dagger}X_{i}-1\right)+\sum_{i}\lambda_{2,i}\left(Y_{i}^{\dagger}Y_{i}-1\right)\nonumber \\
 & +\sum_{i}h_{1,i}+\sum_{i}h_{2,i}.\label{final slave rotor H}
\end{align}

\noindent Here, $P_{i}$ and $Q_{i}$ are the momenta conjugated to
the coordinates $X_{i}$ and $Y_{i}$, respectively. $\mu$ is the
chemical potential of itinerant electrons. $\lambda_{1,i}$ and $h_{1,i}$
are the Lagrange multipliers that ensure the constraints $X_{i}^{\dagger}X_{i}=1$
and $L_{X,i}=i(X_{i}P_{i}-X_{i}^{\dagger}P_{i}^{\dagger})=\sum_{\sigma}f_{i,\sigma}^{\dagger}f_{i,\sigma}-1$
hold. Similarly, $\lambda_{1,i}$ and $h_{1,i}$ regarding $Y_{i}$
fields are defined. We use the Hubbard-Stratonovich transformation
to decompose the four-field terms with auxiliary fields. The Eq. \eqref{final slave rotor H}
then becomes $H=H_{\mathrm{QSL}}+H_{\mathrm{AL}}+H_{\mathrm{c}}$,
where $H_{\mathrm{QSL}}$ describes the host electron layer with

\begin{align}
H_{\mathrm{QSL}} & =\sum_{i,j,\sigma}t_{ij}\chi_{ji}^{X}f_{i,\sigma}^{\dagger}f_{j,\sigma}+\mathrm{h.c.}-\sum_{i,\sigma}\left(\mu+h_{1,i}\right)f_{i,\sigma}^{\dagger}f_{i,\sigma}\nonumber \\
 & +\sum_{i,j}\chi_{ij}^{f}X_{j}^{\dagger}X_{i}+U_{\mathrm{QSL}}\sum_{i}P_{i}^{\dagger}P_{i}\nonumber \\
 & +\sum_{i}\lambda_{1,i}(X_{i}^{\dagger}X_{i}-1)-\sum_{i,j}\chi_{ji}^{X}\chi_{ij}^{f},\label{eq:QSL slave rotor H}
\end{align}
$H_{\mathrm{AL}}$ describes the second layer consisting of Anderson
impurities with 
\begin{align}
H_{\mathrm{AL}}= & \sum_{i,\sigma}\left(\epsilon_{d}-h_{2}\right)a_{i,\sigma}^{\dagger}a_{i,\sigma}+\sum_{i}\lambda_{2,i}\left(Y_{i}^{\dagger}Y_{i}-1\right)+\sum_{i}h_{2,i}\nonumber \\
 & +U\sum_{i}Q_{i}^{\dagger}Q_{i}+i\sum_{i}h_{2,i}Q_{i}Y_{i}-i\sum_{i}h_{2,i}Y_{i}^{\dagger}Q_{i}^{\dagger},\label{eq: QSL slave rotor AL}
\end{align}
and $H_{\mathrm{c}}$ describes the coupling between the two layers
\begin{equation}
H_{\mathrm{c}}=-\sum_{i}u_{i}Y_{i}^{\dagger}X_{i}+\sum_{i,\sigma}w_{i}f_{i,\sigma}^{\dagger}a_{i,\sigma}+\sum_{i}\frac{2u_{i}w_{i}}{V}.\label{Hc}
\end{equation}

In the mean-field approximation, the coupled fields and Lagrange multipliers
satisfy the following self-consistent equations at the saddle point:

\begin{align}
u= & -\frac{2V}{\beta N}\sum_{\boldsymbol{k},n}G(a,f^{\dagger},i\omega_{n},\boldsymbol{k},\sigma),\label{SCE u}\\
w= & -\frac{V}{\beta N}\sum_{\boldsymbol{k},n}G(X,Y^{\dagger},i\nu_{n},\boldsymbol{k}),\label{SCE w}\\
1= & -\frac{1}{\beta N}\sum_{\boldsymbol{k},n}G(Y,Y^{\dagger},i\nu_{n},\boldsymbol{k})e^{i\nu_{n}0^{+}},\label{SCE lambda}\\
0= & -\frac{1}{2U\beta N}\sum_{\boldsymbol{k},n}i\nu_{n}G\left(Y,Y^{\dagger},i\nu_{n},\boldsymbol{k}\right)\left[e^{i\nu_{n}0^{+}}+e^{-i\nu_{n}0^{+}}\right]\nonumber \\
 & +\frac{h_{2}}{U}+\frac{1}{\beta N}\sum_{\boldsymbol{k},n}G\left(a,a^{\dagger},i\omega_{n},\boldsymbol{k},\sigma\right)-\frac{1}{2}.\label{SCE h2}
\end{align}

\begin{figure}[t]
\center 
\includegraphics[width=1\linewidth]{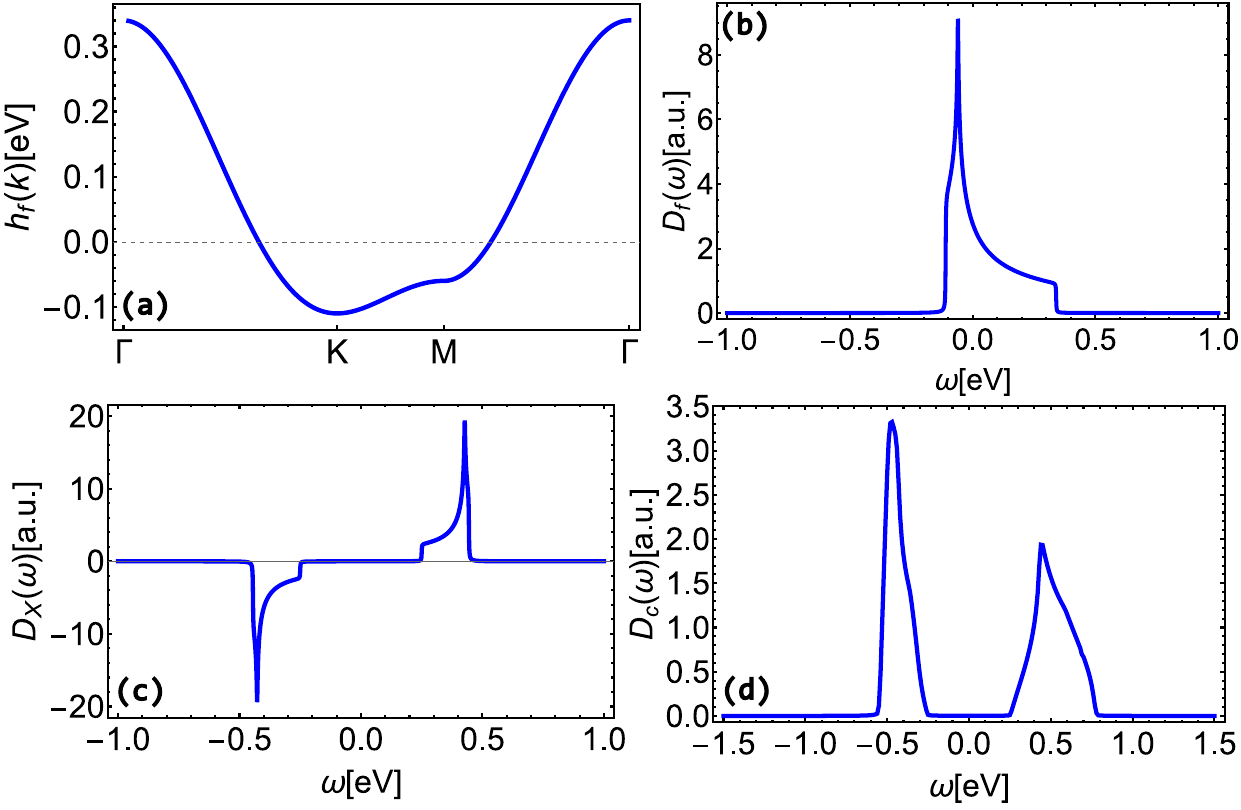} 
\caption{(a) Spinon dispersion relation of isolated U(1) spin liquid, along
path through symmetry points $\Gamma,\mathrm{K},\mathrm{M},\Gamma$
in triangular lattice Brillouin Zone. (b), (c) The spectral functions
of spinons and chargons of U(1) spin liquid respectively. The spectral
functions is obtained from $D(\omega)=-\frac{1}{\pi N}\sum_{\boldsymbol{k}}\mathrm{Im}G(\omega+i0^{+},\boldsymbol{k})$.
(d) Electron spectral function, which is obtained by convolution of
spinons and chargons' Green's functions.}
\label{Fig-iso_qsl} 
\end{figure}

\noindent In these equations, we set $h_{2,i}=0$, since $L_{Y,i}=\sum_{\sigma}a_{i,\sigma}^{\dagger}a_{i,\sigma}-1=0$
is always a solution of Eq. \eqref{SCE h2}, which means that the
impurity lattice still maintains the single-occupation state of the
electrons. The constant $N$ is the number of unit cells and $\beta=1/k_{B}T$
is the inverse temperature. $\omega_{n}=(2n+1)\pi/\beta$ and $\nu_{n}=2n\pi/\beta$
are the fermionic and bosonic Matsubara frequencies, respectively.
Additionally, we have used the spin liquid mean-field Hamiltonian
that matches the recent experiment on single-layer 1T-TaSe$_{2}$
\citep{Chen2022}:

\begin{align}
H_{\mathrm{QSL}}=\sum_{\boldsymbol{k},\sigma}h_{f}(\boldsymbol{k})f_{\boldsymbol{k},\sigma}^{\dagger}f_{\boldsymbol{k},\sigma}+\sum_{k}\omega_{X}^{2}(\boldsymbol{k})X_{\boldsymbol{k}}^{\dagger}X_{\boldsymbol{k}}+U_{\mathrm{QSL}}\sum_{\boldsymbol{k}}P_{\boldsymbol{k}}^{\dagger}P_{\boldsymbol{k}},\label{QSL H}
\end{align}

\noindent where spinon energy $h_{f}(\boldsymbol{k})=t_{F}\gamma(\boldsymbol{k})-\mu$,
$\omega_{X}(\boldsymbol{k})=\sqrt{-t_{X}\gamma(\boldsymbol{k})+\lambda_{1}}$
is chargon frequency, and nearest neighbor form factor $\gamma(\boldsymbol{k})=2(2\cos\frac{1}{2}k_{x}a\cos\frac{\sqrt{3}}{2}k_{y}a+\cos k_{x}a)$
and $a$ is lattice constant. Here, $t_{F}=0.05~\mathrm{eV}$, $t_{X}=0.019~\mathrm{eV}$
are spinon and chargon hopping, $\mu=-0.04~\mathrm{eV}$ is the spinon
chemical potential, and local interaction $U_{\mathrm{QSL}}=0.775~\mathrm{eV}$,
$\lambda_{1}=0.157~\mathrm{eV}$ is Lagrange multiplier. The relation
between Mott gap and parameter $\lambda_{1}$ are $\Delta_{g}=\sqrt{U_{\mathrm{QSL}}(\lambda_{1}-6t_{F})}=0.25~\mathrm{eV}$.
The Green's functions in equations above are (see Appendix \ref{Appendix B}
for details):

\begin{equation}
G^{0}(f,f^{\dagger},i\omega_{n},\boldsymbol{k},\sigma)=\frac{1}{i\omega_{n}-h_{f}(\boldsymbol{k})},\label{GFff0}
\end{equation}
\begin{equation}
G^{0}(a,a^{\dagger},i\omega_{n},\sigma)=\frac{1}{i\omega_{n}-\epsilon_{d}},\label{GFaa0}
\end{equation}
\begin{equation}
G(f,a^{\dagger},i\omega_{n},\boldsymbol{k},\sigma)=\frac{wVG^{0}(f,f^{\dagger},i\omega_{n},\boldsymbol{k},\sigma)}{i\omega_{n}-\epsilon_{0}-w^{2}G^{0}(f,f^{\dagger},i\omega_{n},\boldsymbol{k},\sigma)},\label{GFaf}
\end{equation}
\begin{equation}
G^{0}(X,X^{\dagger},i\nu_{n},\boldsymbol{k})=\frac{-1}{\frac{\nu_{n}^{2}}{U_{QSL}}+\omega_{X}^{2}(\boldsymbol{k})},\label{GFXX0}
\end{equation}
\begin{equation}
G(X,Y^{\dagger},i\nu_{n},\boldsymbol{k})=\frac{uG^{0}(X,X^{\dagger},i\omega_{n},\boldsymbol{k})}{\frac{\nu_{n}^{2}}{U}+\lambda_{2}+u^{2}G^{0}(X,X^{\dagger},i\nu_{n},\boldsymbol{k})},\label{GFXY}
\end{equation}
\begin{equation}
G(Y,Y^{\dagger},i\nu_{n},\boldsymbol{k})=\frac{-1}{\frac{\nu_{n}^{2}}{U}+\lambda_{2}+u^{2}G^{0}(X,X^{\dagger},i\nu_{n},\boldsymbol{k})}.\label{GFYY}
\end{equation}

\subsection{Uncoupled model: the spin liquid phase}

First, let us set $V=0$, the uncoupled layers. As pointed out earlier,
we consider the limit of a large $U_{\mathrm{QSL}}/t$ ensuring that
the system is deep within the Mott insulator phase with a spin liquid
ground state. Within the slave-rotor framework, the insulating Mott
phase is characterized by the vanishing quasiparticle weight given
by the expectation value of the rotor field $Z=\langle X\rangle$
\cite{Florens2002,Florens2004,Senthil2008,Podolsky2009}, implying
that the charge is stripped off electrons. In Fig.~\ref{Fig-iso_qsl},
we show the electronic structure of the Mott phase. Fig.~\ref{Fig-iso_qsl}(a)
depicts the energy band dispersion of the spinons $h_{f}(\boldsymbol{k})$
on the triangular lattice. The Fermi level corresponding to half-filling
is shown by a dashed line, and it is seen that the spinons form a
Fermi surface. Fig.~\ref{Fig-iso_qsl}(b) shows the corresponding
density of states of spinons $D_{f}(\omega)=(1/N)\sum_{\boldsymbol{k}}A_{f}(\omega,\boldsymbol{k})$
where $A_{f}(\omega,\boldsymbol{k})=-(1/\pi)\mathrm{Im}G^{0}(f,f^{\dagger},\omega+i0^{+},\boldsymbol{k},\sigma)$
is the spinon spectral density. The density of states of chargons
$D_{X}(\omega)=(1/N)\sum_{\boldsymbol{k}}A_{X}(\omega,\boldsymbol{k})$
with $A_{X}(\omega,\boldsymbol{k})=-(1/\pi)\mathrm{Im}G^{0}(X,X^{\dagger},\omega+i0^{+},\boldsymbol{k})$
is shown in Fig.~\ref{Fig-iso_qsl}(c). The Mottness of the original
electrons is, however, given by the spectral density of the convoluted
spinon and chargon Green's functions, $D_{c}(\omega)=(1/N)\sum_{\boldsymbol{k}}A_{c}(\omega,\boldsymbol{k})$
with $A_{c}(\omega,\boldsymbol{k})=-(1/\pi)\mathrm{Im}G^{0}(c,c^{\dagger},\omega+i0^{+},\boldsymbol{k},\sigma)$,
where $G^{0}(c,c^{\dagger},i\omega_{n},\boldsymbol{k},\sigma)=\beta^{-1}\sum_{\nu_{m}}G^{0}(f,f^{\dagger},i\omega_{n}+i\nu_{m},\boldsymbol{k},\sigma)G^{0}(X,X^{\dagger},i\nu_{m},\boldsymbol{k})$.
The density of states $D(\omega)$ is shown in Fig.~\ref{Fig-iso_qsl}(d),
where the formation of upper and lower Hubbard bands is clearly seen.

\begin{figure}[t]
\center 
\includegraphics[width=1\linewidth]{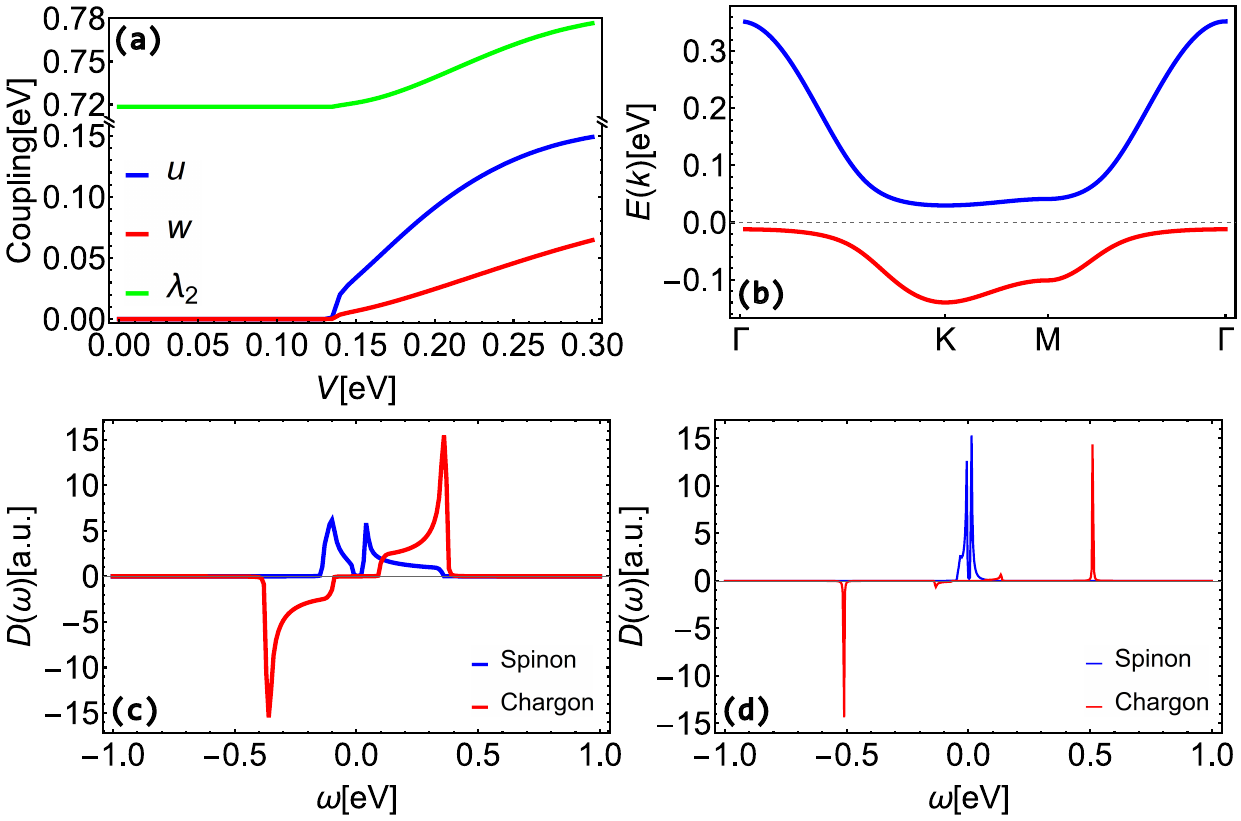} 
\caption{(a) The mean field parameters $u$ and $w$ as a function of the coupling
$V$. (b) The spinon dispersion relation in spinon Kondo lattice phase.
(c) Spinon and chargon spectral functions of spin liquid part in spinon
Kondo lattice phase. (d) Spinon and chargon spectral functions of
Anderson lattice part in spinon Kondo lattice phase.}
\label{Fig phase diagram and dos} 
\end{figure}

\subsection{Spinon Kondo lattice phase}

\phantom{}

Having established the spin liquid phase on the triangular lattice
as described in the preceding subsection, we now consider the hybridization
of the spin liquid phase with a lattice of Anderson impurities. The
coupling strength is given by $V\neq0$ (see Eq. \eqref{original H}),
whose effects are encapsulated in the fields $u_{i}$ and $w_{i}$
in Eq. \eqref{Hc}, which determine the hybridization between the
spinons (chargons) of the spin liquid phase and the spinons (chargons)
on the Anderson impurity, respectively. To examine the effects of
$V\neq0$, we solved the self-consistent equations in Eqs. \eqref{SCE u}-\eqref{SCE w}
along with the constraints in \eqref{SCE lambda} and \eqref{SCE h2}
numerically.

The variation of the hybridization fields $u$ and $w$ as a function
of the coupling strength is shown in Fig.~\ref{Fig phase diagram and dos}(a).
There is a critical coupling strength $V_{c}$ beyond which the hybridization
fields $u$ and $w$ acquire nonzero values. When $V<V_{c}$, $u$
and $w$ are equal to zero, indicating that the system consists of
two separate layers; the spin liquid phase and the Anderson impurity
lattice are uncoupled. For $V>V_{c}$, $u$ and $w$ are greater than
zero, placing the system in the hybridized phase characterized by
heavy spinons near the Fermi level, where the dispersion becomes nearly
flat, as seen in Fig.~\ref{Fig phase diagram and dos}(b). This phase
is termed a spinon Kondo lattice phase. It is important to note the
distinct difference between the spinon Kondo lattice phase and the
normal heavy fermions in the Kondo lattice model. In the latter, a
normal metal is antiferromagnetically coupled to a lattice of magnetic
impurities as $J\sum_{i}c_{i}^{\dagger}\boldsymbol{\sigma}c_{i}\cdot\boldsymbol{S}_{i}$.
Here, the coupling strength $J$ is relevant, and the model transitions
to a heavy fermion model as soon as $J\neq0$. However, in our model,
there is a critical value of coupling strength, $V_{c}$, where the
phase transition occurs.

Furthermore, the spinon Kondo lattice phase is characterized by two
spinon bands separated by a small gap, as shown in Fig.~\ref{Fig phase diagram and dos}(b),
analogous to the Kondo insulator phase. Indeed, the hybridized bands
open a gap between them. Since the spin liquid phase and the impurity
lattice are both singly occupied, the spinon occupation number (1+1)
mod 2 is equal to zero, leading the system to form a spinon Kondo
insulator.

\section{Spinon-chargon bound states\label{sec III}}

\begin{figure}
\centering{}\includegraphics[width=1\linewidth]{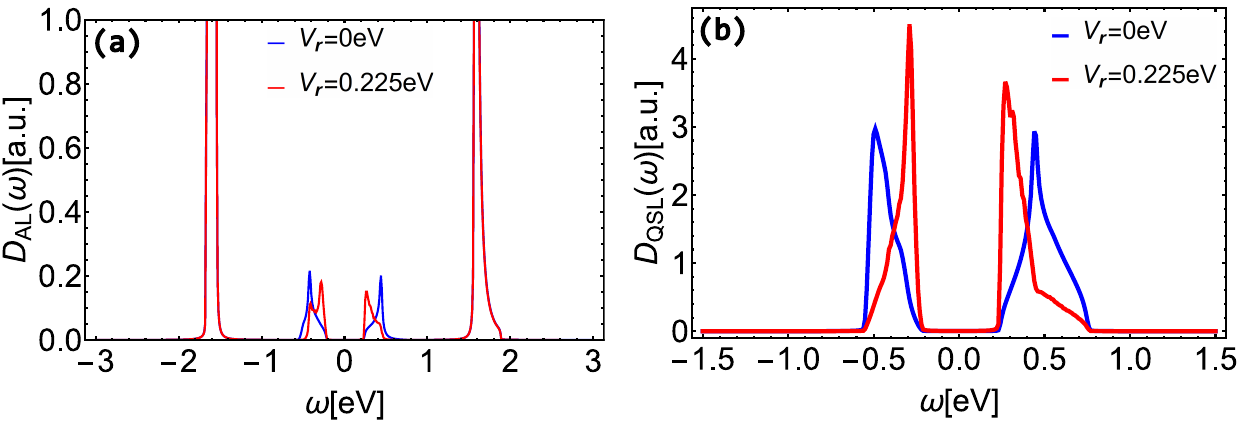}\caption{(a) Spectral function of electron for Anderson lattice part in spinon
Kondo lattice phase, where blue and red line refer the interaction
strength $V_{\bm{r}}=0\ \mathrm{eV}$ and $V_{\bm{r}}=0.225\ \mathrm{eV}$
respectively. (b) Spectral function of electron for QSL part in spinon
Kondo lattice phase, where blue and red line refer the same meaning
as (a).}
\label{Electron dos with correction} 
\end{figure}

In a U(1) quantum spin liquid, there exists an emergent U(1) gauge
symmetry between the spinons and chargons. At low energies, the corresponding
U(1) gauge field is noncompact and mediates a Coulomb potential \cite{Hermele2004,Lee2009}
with the assumption of deconfined quantum spin liquid, which tends to bind the spinons
and chargons together into electrons \citep{Lee1998,Lee2006,Lee2009,He2022}.To
analyze bound states, we employ the Bethe-Salpeter equation to calculate
the electron Green's functions. We define new field operators $\psi_{s}(i\omega_{n},\boldsymbol{k},\sigma)=(a(i\omega_{n},\sigma),f(i\omega_{n},\boldsymbol{k},\sigma))^{T}$,
$Z_{c}(i\nu_{n},\boldsymbol{k})=(Y(i\nu_{n}),X(i\nu_{n},\boldsymbol{k}))^{T}$
for spinons and chargons, respectively.

In the ladder approximation, the Bethe-Salpeter equation is given
by \citep{Salpeter1951,Greiner2003,Lurie1968}:

\begin{align}
G_{\psi_{e}}(k_{1})= & -(\beta N)^{-1}\sum_{q}G_{\psi_{s}}(k+q)\otimes G_{Z_{c}}(q)\nonumber \\
 & \times\left(\mathbf{1}_{2}-(\beta N)^{-1}\sum_{q}K_{p}^{*}\left(i\nu_{n}\right)G_{\psi_{s}}(k+q)\otimes G_{Z_{c}}(q)\right)^{-1},\label{electron GF}
\end{align}

\noindent where $k=(i\omega_{n},\boldsymbol{k})$, $q=(i\nu_{n},\boldsymbol{q})$,
the Green's functions $G_{\psi_{e/s}}(k)$ and $G_{Z_{c}}(q)$ are
defined as $G_{\psi_{e/s}}(k)=G(\psi_{e/s},\psi_{e/s}^{\dagger},i\omega_{n},\boldsymbol{k},\sigma)$,
and ( $G_{Z_{c}}(q)=G(Z_{c},Z_{c}^{\dagger},i\nu_{n},\boldsymbol{q})$,
respectively. In this context, $\otimes$ denotes the Kronecker product,
resulting in a $4\times4$ matrix, and we consider only the $i,j\in\{1,4\}$
block. $\mathbf{1}_{2}$ represents the two-dimensional identity matrix.
The $2\times2$ matrix $K^{*}(i\nu_{n})$ is the approximate two-body
interaction kernel with zero entries except for $[K^{*}(i\nu_{n})]_{22}=-\frac{i\nu_{n}}{U_{QSL}}V_{\bm{r}}$.
We focus on the screened Coulomb potential at the same lattice site,
where $V_{\bm{r}}=\varLambda_{f}$ , as detailed in Appendix \ref{Appendix E:Calculate-the-Green's}.
Here, $\varLambda_{f}$ is defined as the spinon half bandwidth, $\varLambda_{f}=(\max[h_{f}(\boldsymbol{k})]-\min[h_{f}(\boldsymbol{k})])/2$.

The electron Green's functions for the Anderson lattice and the spin
liquid part are denoted as $G_{\mathrm{AL}}$ and $G_{\mathrm{QSL}}$,
respectively, and are given by $G_{\mathrm{AL/QSL}}=\left[G(\psi_{e},\psi_{e}^{\dagger},\omega+i0^{+},\boldsymbol{k}_{1},\sigma)\right]_{11/22}$.
Subsequently, we derive their spectral densities $A_{\mathrm{AL/QSL}}(\omega,\boldsymbol{k}_{1})=-\frac{1}{\pi}\mathrm{Im}\left[G(\psi_{e},\psi_{e}^{\dagger},\omega+i0^{+},\boldsymbol{k}_{1},\sigma)\right]_{11/22}$
and the density of states for each lattice as $D_{\mathrm{AL/QSL}}(\omega)=-\frac{1}{\pi N}\sum_{\boldsymbol{q},\sigma}A_{\mathrm{AL/QSL}}(\omega,\boldsymbol{k}_{1})$.
The results are illustrated in Fig.~\ref{Electron dos with correction}.
Notably, the very sharp peaks near $\omega\simeq1.8\ \mathrm{eV}$
correspond to the lower and upper Hubbard excitations on the Anderson
impurity, which remain unaffected by the Coulomb potential. However,
the middle bands, which arise from hybridization with the spin liquid,
are influenced by the formation of the bound state. The impact of
the Coulomb potential and the spinon-chargon bound states is more
pronounced in the density of states of the parent quantum spin liquid.
As depicted in Fig.~\ref{Electron dos with correction}(b), the Coulomb
potential shifts the correlated excitations at high energies toward
the edge of the Hubbard band.

\section{Theromodynmic properties with gauge field correction\label{sec IV}}

\subsection{Neutron scattering: spinon susceptibility}

Neutron scattering can measure the collective excitations of a system,
and the response is characterized by the spinon susceptibility. In
a spin liquid with gapless spinons, emergent U(1) gauge fields mediate
a Coulomb potential $V(\boldsymbol{r}-\boldsymbol{r}^{\prime})=\frac{g^{2}}{\left|\boldsymbol{r}-\boldsymbol{r}^{\prime}\right|}$
between the spinons (refer to Appendix \ref{sec:Emergent--gaugeAppendix D: Emergent U(1) gauge field in quantum spin liquid}
for detailed derivations). Consequently, collective excitations are
anticipated to manifest in the higher energy regions and may be detectable
in neutron scattering experiments on quantum spin liquid states \citep{Banerjee2016}.
In our spinon Kondo lattice phase, the emergent Coulomb potential
causes the magnetic excitations to exhibit significant variations
compared to the non-interacting case.

We analyze the longitudinal component of the magnetic susceptibility
$\chi(i\nu_{n},\bm{q})=-\int_{0}^{\beta}\left\langle S_{z}(\tau,\bm{q})S_{z}(0,\bm{q})\right\rangle e^{i\nu_{n}\tau}d\tau$,
where $S_{z}(\tau,\bm{q})$ denotes the $z$-component of the electron
spin operator $S_{z}(\tau,\bm{q})=c_{\boldsymbol{q},\uparrow}^{\dagger}(\tau)c_{\boldsymbol{q},\uparrow}(\tau)-c_{\boldsymbol{q},\downarrow}^{\dagger}(\tau)c_{\boldsymbol{q},\downarrow}(\tau)$.
Given that the chargons in the spin liquid possess an energy gap,
their contribution to the ground state is negligible; therefore, we
focus solely on calculating the spinon susceptibility. The expression
for the non-interacting spinon susceptibility is as follows:

\begin{align}
\chi_{0}(i\nu_{n},\boldsymbol{q})=\frac{1}{\beta N}\sum_{k,\sigma}G(k+q,\sigma)\otimes G(k,\sigma),
\end{align}
where $G(k,\sigma)=G(\psi_{s},\psi_{s}^{\dagger},i\omega_{n},\boldsymbol{k},\sigma)$.
The random-phase approximation (RPA) susceptibility is 
\begin{equation}
\chi_{\mathrm{RPA}}(i\nu_{n},\boldsymbol{q})=\frac{\chi_{0}(i\nu_{n},\boldsymbol{q})}{1-V(\boldsymbol{q})\chi_{0}(i\nu_{n},\boldsymbol{q})},
\end{equation}
from which we calculate the spectral density $A_{\mathrm{RPA}}=-\frac{1}{\pi}\mathrm{Im}\chi_{\mathrm{RPA}}$.

\begin{figure}
\noindent \begin{centering}
\includegraphics[width=3.5in]{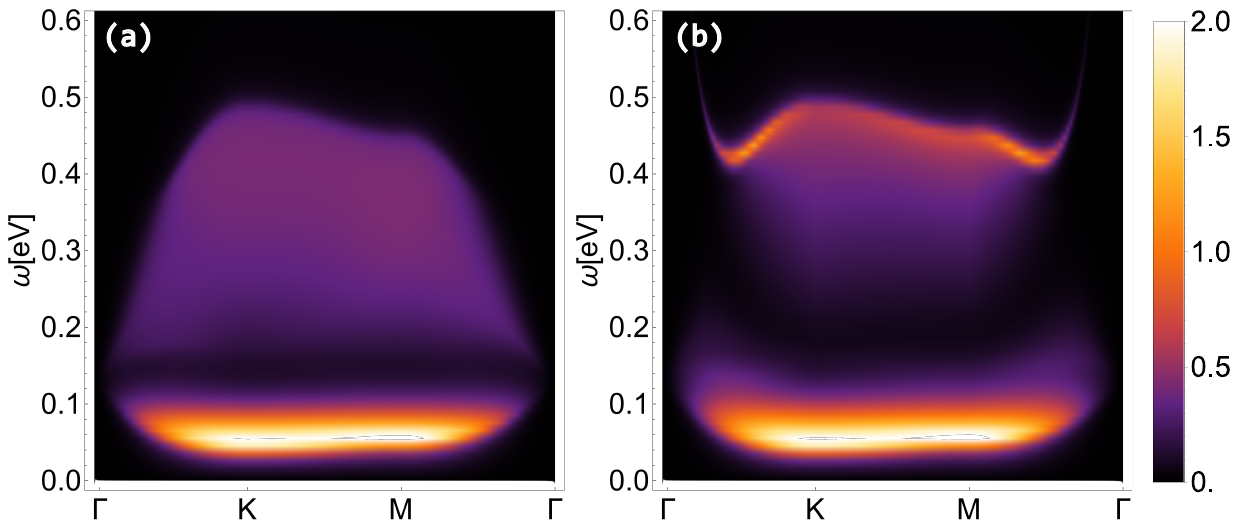} 
\par\end{centering}
\caption{(a),(b) Spectral density of spinon susceptibility $\chi_{0/\mathrm{RPA}}$
in spinon Kondo lattice phase along high symmetry points without and
with RPA correction.}

\label{spinon susceptibility} 
\end{figure}

Fig.~\ref{spinon susceptibility} (a) and (b) depict the excitation
spectrum of the spinon Kondo lattice phase, considering the bare and
RPA susceptibilities, respectively. A low-energy branch of excitations
is present, along with a continuum of particle-hole excitations at
higher energies within the bare spectral density. The coupling between
the parent quantum spin liquid and the Anderson impurity lattice is
evidenced by a gap in the excitation spectrum. Fig.~\ref{spinon susceptibility}(b)
presents the same excitation spectrum while incorporating the Coulomb
interaction $V(\boldsymbol{q})$ through the RPA. The low-energy branch
remains largely unaffected, but the upper continuum undergoes significant
modifications due to the Coulomb interaction. Notably, the Coulomb
interaction propels the magnetic excitations near the $\Gamma=(0,0)$
point to substantially higher energies.

\subsection{Internal energy and specific heat}

We analyze the internal energy and specific heat of the spinon Kondo
lattice phase using the bound-state electron Green's function. For
a U(1) quantum spin liquid with a spinon Fermi surface, the specific
heat and thermal conductivity resemble those of a typical Fermi liquid;
that is, they are proportional to the temperature at low temperatures
\citep{Ribak2017,Murayama2020}. However, in our spinon Kondo lattice
phase, the presence of the Anderson lattice leads to the formation
of a spinon Kondo insulator, as depicted in Fig.~\ref{Fig phase diagram and dos}(b),
indicating that its internal energy and specific heat properties should
differ.

\begin{figure}
\noindent \begin{centering}
\includegraphics[width=3.5in]{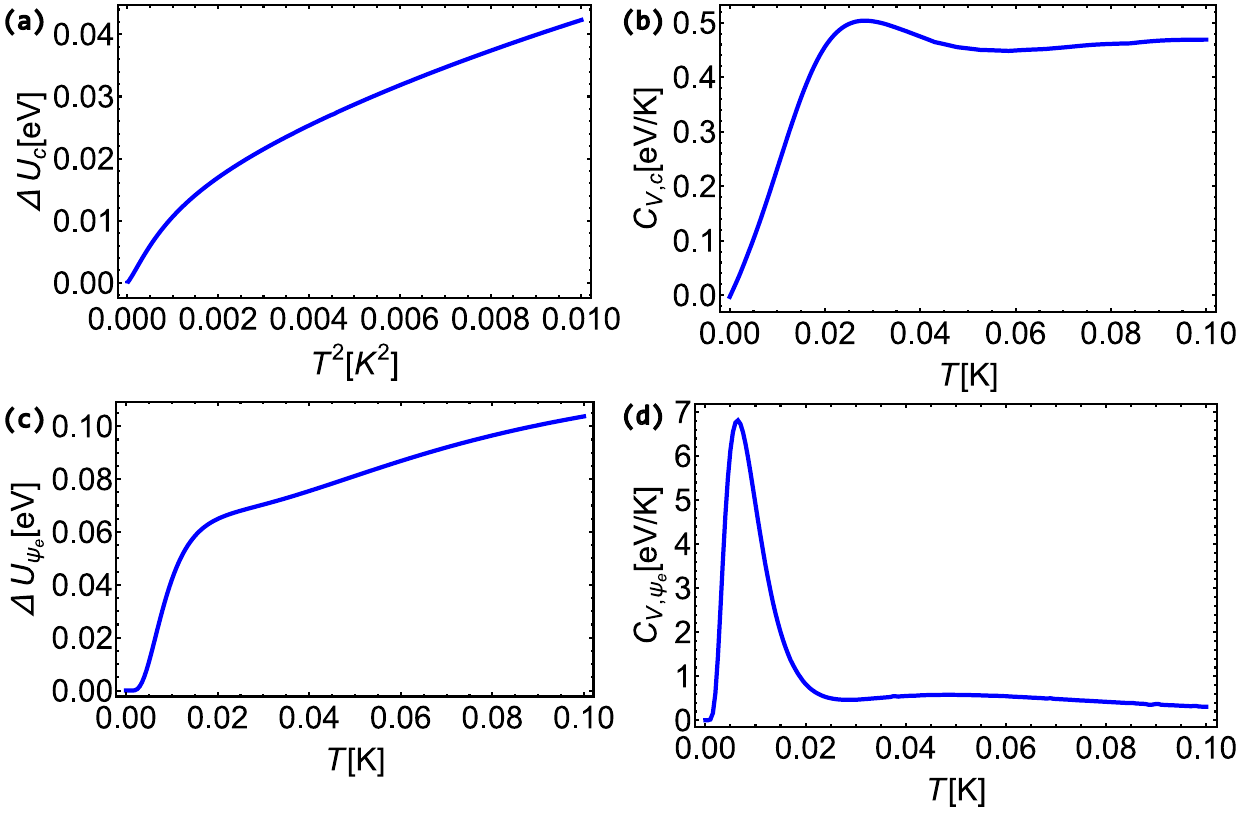} 
\par\end{centering}
\caption{(a) Internal energy difference $\Delta U_{c}=U_{c}-U_{c,0}$ of electrons
in QSL respect to temperature square $T^{2}$, where $U_{c}$ QSL
electron internal energy and the $U_{c,0}$ is defined as the internal
energy of QSL electron at $T=0\ \mathrm{K}$. (b) Temperature dependence
of specific heat of QSL electron. (c) Internal energy difference $\Delta U_{\psi_{e}}$
of electron in spinon Kondo lattice phase respect to temperature $T$.
$U_{\psi_{e}}$ and $U_{\psi_{e},0}$ are the electron internal in
spinon Kondo lattice phase in arbitrary temperature and $T=0\ \mathrm{K}$
respectively. (d) Temperature dependence of electron specific heat
in spinon Kondo lattice phase.}

\label{IEandSH} 
\end{figure}

The internal energy is calculated using the formula $U=\left\langle H\right\rangle =\int\omega D(\omega)n_{f}(\omega)d\omega$,
where $D(\omega)$ is the spectral function. As discussed below Eq.
\eqref{electron GF}, the electron density for the spinon Kondo lattice
phase is $D_{\psi_{e}}(\omega)=-\frac{1}{\pi N}\sum_{\boldsymbol{q},\sigma,i,j}\mathrm{Im}\left[G(\psi_{e},\psi_{e}^{\dagger},\omega+i0^{+},\boldsymbol{q},\sigma)\right]_{ij}$.
For comparison with a quantum spin liquid phase, we also consider
the density of states of the latter phase as $D_{c}(\omega)=-\frac{1}{\pi N}\sum_{\boldsymbol{q},\sigma}\mathrm{Im}G(c,c^{\dagger},\omega+i0^{+},\boldsymbol{q},\sigma)$;
for details see Appendix \ref{Appendix E:Calculate-the-Green's}.
Here, $n_{f}(\omega)=\frac{1}{\exp\left(\beta\omega\right)+1}$ represents
the Fermi distribution function. With these, the internal energy and
specific heat can be calculated, and the results are shown in Fig.~\ref{IEandSH}.

As demonstrated in Fig.~\ref{IEandSH}(a), the internal energy of
the pure spin liquid exhibits a linear relationship with the square
of the temperature at low temperatures. This results in the specific
heat being proportional to the temperature, as anticipated in the
low-temperature regime approaching zero, as depicted in Fig.~\ref{IEandSH}(b).
In contrast, the electron specific heat of the spinon Kondo lattice
phase remains invariant at low temperatures. This leads to a pronounced
peak in the specific heat at low temperatures following a minimal
zero plateau near $T=0\ \mathrm{K}$, indicative of insulator-like
behavior.

\section{CONCLUSIONs\label{conclusions}}

This work is primarily inspired by the recent experimental observations
of the Kondo resonant state in cobalt atoms on a single-layer 1T-TaSe$_{2}$,
which is a Mott insulator with a spin liquid ground state \citep{Chen2022}.
This phenomenon is akin to the Kondo effects observed in normal metals
doped with dilute magnetic impurities. Theoretically, we consider
a lattice of single-level quantum dots, the Anderson impurity lattice,
which is coupled to a U(1) spin liquid with spinon Fermi surface.
To address the Mott insulator and spin liquid phases, we employ the
slave-rotor parton construction for both the Anderson lattice and
the parent quantum spin liquid layer.

Let us briefly recapitulate the main findings and answers to questions
posed in the introduction: (i) In the regime of strong Hubbard interaction
across all lattice sites, including the parent triangular and Anderson
impurity lattices, spinons persist as the sole low-energy degrees
of freedom. Beyond a critical coupling between the parent triangular
and Anderson impurity lattices leads to the hybridization of spinons
across different layers, effectively dissolving localized spinons
into the spinon Fermi surface of the spin liquid state. (ii) The interaction
between the Anderson impurity lattice and the U(1) spin liquid gives
rise to a novel phase, termed the spinon Kondo lattice phase. Hybridization
of spinons from both lattices results in the formation of two spinon
energy bands. Owing to half-filling of both lattices, the lower band
is fully occupied and separated by a minor energy gap from the upper
band, rendering the original spinon Fermi surface fully gapped and
the hybridized phase a spinon Kondo insulator. (iii) The slave-rotor
method inherently allows for an emergent local U(1) gauge symmetry.
We investigated the influence of U(1) gauge fields by computing the
bound state between spinons and chargons. Our many-body calculations
indicate that bound state fluctuations do not significantly alter
the spinon spectrum, only shifting high-energy states towards the
proximate Mott band edges. (iv) Lastly, we examined the potential
response of our model to neutron scattering measurements by assessing
the magnetic susceptibility and to thermal measurements by evaluating
the specific heat. Both measurements exhibit characteristics indicative
of the spinon Kondo insulator phase.

Following up our work presented here, there are a few directions that we 
leave for future studies. In one front, one may consider the tunneling 
between the quantum dots and explore how it may affect the spinon Kondo lattice phase. 
This might parallel the spinon analogue of the
FL$^{*}$ phase. On another front, a natural extension of our study
would be to consider broader classes of non-Abelian spin liquids with
Fermi surfaces, such as the SU(2) spin liquid \citep{Lee1998,Lee2006,Hermele2007,Chen2012},
and their coupling to magnetic impurities. The presence of spinon
pairing terms could lead to the discovery of various exotic topological
quantum phases, including non-Abelian topological order, non-Abelian
spinon metals, etc., which may serve as potential models for topological
quantum computation \citep{Kitaev2003,Nayak2008}. Thus, the spinon
Kondo lattice phase characterized in this study presents potential
theoretical and experimental framework for the investigation of novel
phases.

\section*{ACKNOWLEDGMENTS}

The authors would like to thank Sharif University of Technology for
supports. ZXM especially thanks W.-Y. He for helpful discussion on
programming the single-impuity case. ZXM also thanks HongLiang Wei
for constructive discussions on physics and programming. MK also would
like to thank INSF-Grant No. 4027770.


\begin{thebibliography}{40}%
\makeatletter
\providecommand \@ifxundefined [1]{%
 \@ifx{#1\undefined}
}%
\providecommand \@ifnum [1]{%
 \ifnum #1\expandafter \@firstoftwo
 \else \expandafter \@secondoftwo
 \fi
}%
\providecommand \@ifx [1]{%
 \ifx #1\expandafter \@firstoftwo
 \else \expandafter \@secondoftwo
 \fi
}%
\providecommand \natexlab [1]{#1}%
\providecommand \enquote  [1]{``#1''}%
\providecommand \bibnamefont  [1]{#1}%
\providecommand \bibfnamefont [1]{#1}%
\providecommand \citenamefont [1]{#1}%
\providecommand \href@noop [0]{\@secondoftwo}%
\providecommand \href [0]{\begingroup \@sanitize@url \@href}%
\providecommand \@href[1]{\@@startlink{#1}\@@href}%
\providecommand \@@href[1]{\endgroup#1\@@endlink}%
\providecommand \@sanitize@url [0]{\catcode `\\12\catcode `\$12\catcode
  `\&12\catcode `\#12\catcode `\^12\catcode `\_12\catcode `\%12\relax}%
\providecommand \@@startlink[1]{}%
\providecommand \@@endlink[0]{}%
\providecommand \url  [0]{\begingroup\@sanitize@url \@url }%
\providecommand \@url [1]{\endgroup\@href {#1}{\urlprefix }}%
\providecommand \urlprefix  [0]{URL }%
\providecommand \Eprint [0]{\href }%
\providecommand \doibase [0]{http://dx.doi.org/}%
\providecommand \selectlanguage [0]{\@gobble}%
\providecommand \bibinfo  [0]{\@secondoftwo}%
\providecommand \bibfield  [0]{\@secondoftwo}%
\providecommand \translation [1]{[#1]}%
\providecommand \BibitemOpen [0]{}%
\providecommand \bibitemStop [0]{}%
\providecommand \bibitemNoStop [0]{.\EOS\space}%
\providecommand \EOS [0]{\spacefactor3000\relax}%
\providecommand \BibitemShut  [1]{\csname bibitem#1\endcsname}%
\let\auto@bib@innerbib\@empty
\bibitem [{\citenamefont {Anderson}(1973)}]{Anderson1973}%
  \BibitemOpen
  \bibfield  {author} {\bibinfo {author} {\bibfnamefont {P.}~\bibnamefont
  {Anderson}},\ }\href {\doibase https://doi.org/10.1016/0025-5408(73)90167-0}
  {\bibfield  {journal} {\bibinfo  {journal} {Materials Research Bulletin}\
  }\textbf {\bibinfo {volume} {8}},\ \bibinfo {pages} {153} (\bibinfo {year}
  {1973})}\BibitemShut {NoStop}%
\bibitem [{\citenamefont {Anderson}(1987)}]{Anderson1987}%
  \BibitemOpen
  \bibfield  {author} {\bibinfo {author} {\bibfnamefont {P.~W.}\ \bibnamefont
  {Anderson}},\ }\href {\doibase 10.1126/science.235.4793.1196} {\bibfield
  {journal} {\bibinfo  {journal} {Science}\ }\textbf {\bibinfo {volume}
  {235}},\ \bibinfo {pages} {1196} (\bibinfo {year} {1987})}\BibitemShut
  {NoStop}%
\bibitem [{\citenamefont {Wen}(2002)}]{Wen2002}%
  \BibitemOpen
  \bibfield  {author} {\bibinfo {author} {\bibfnamefont {X.-G.}\ \bibnamefont
  {Wen}},\ }\href {\doibase 10.1103/PhysRevB.65.165113} {\bibfield  {journal}
  {\bibinfo  {journal} {Phys. Rev. B}\ }\textbf {\bibinfo {volume} {65}},\
  \bibinfo {pages} {165113} (\bibinfo {year} {2002})}\BibitemShut {NoStop}%
\bibitem [{\citenamefont {Wen}(2004)}]{Wen2004}%
  \BibitemOpen
  \bibfield  {author} {\bibinfo {author} {\bibfnamefont {X.}~\bibnamefont
  {Wen}},\ }\href {https://books.google.com/books?id=llnlrfdR4YgC} {\emph
  {\bibinfo {title} {Quantum Field Theory of Many-Body Systems: From the Origin
  of Sound to an Origin of Light and Electrons}}},\ Oxford Graduate Texts\
  (\bibinfo  {publisher} {OUP Oxford},\ \bibinfo {year} {2004})\BibitemShut
  {NoStop}%
\bibitem [{\citenamefont {Chen}\ \emph {et~al.}(2010)\citenamefont {Chen},
  \citenamefont {Gu},\ and\ \citenamefont {Wen}}]{Chen2010}%
  \BibitemOpen
  \bibfield  {author} {\bibinfo {author} {\bibfnamefont {X.}~\bibnamefont
  {Chen}}, \bibinfo {author} {\bibfnamefont {Z.-C.}\ \bibnamefont {Gu}}, \ and\
  \bibinfo {author} {\bibfnamefont {X.-G.}\ \bibnamefont {Wen}},\ }\href
  {\doibase 10.1103/PhysRevB.82.155138} {\bibfield  {journal} {\bibinfo
  {journal} {Phys. Rev. B}\ }\textbf {\bibinfo {volume} {82}},\ \bibinfo
  {pages} {155138} (\bibinfo {year} {2010})}\BibitemShut {NoStop}%
\bibitem [{\citenamefont {Lee}\ and\ \citenamefont {Lee}(2005)}]{Lee2005}%
  \BibitemOpen
  \bibfield  {author} {\bibinfo {author} {\bibfnamefont {S.-S.}\ \bibnamefont
  {Lee}}\ and\ \bibinfo {author} {\bibfnamefont {P.~A.}\ \bibnamefont {Lee}},\
  }\href {\doibase 10.1103/PhysRevLett.95.036403} {\bibfield  {journal}
  {\bibinfo  {journal} {Phys. Rev. Lett.}\ }\textbf {\bibinfo {volume} {95}},\
  \bibinfo {pages} {036403} (\bibinfo {year} {2005})}\BibitemShut {NoStop}%
\bibitem [{\citenamefont {Lee}\ \emph {et~al.}(2006)\citenamefont {Lee},
  \citenamefont {Nagaosa},\ and\ \citenamefont {Wen}}]{Lee2006}%
  \BibitemOpen
  \bibfield  {author} {\bibinfo {author} {\bibfnamefont {P.~A.}\ \bibnamefont
  {Lee}}, \bibinfo {author} {\bibfnamefont {N.}~\bibnamefont {Nagaosa}}, \ and\
  \bibinfo {author} {\bibfnamefont {X.-G.}\ \bibnamefont {Wen}},\ }\href
  {\doibase 10.1103/RevModPhys.78.17} {\bibfield  {journal} {\bibinfo
  {journal} {Rev. Mod. Phys.}\ }\textbf {\bibinfo {volume} {78}},\ \bibinfo
  {pages} {17} (\bibinfo {year} {2006})}\BibitemShut {NoStop}%
\bibitem [{\citenamefont {Lee}\ \emph {et~al.}(1998)\citenamefont {Lee},
  \citenamefont {Nagaosa}, \citenamefont {Ng},\ and\ \citenamefont
  {Wen}}]{Lee1998}%
  \BibitemOpen
  \bibfield  {author} {\bibinfo {author} {\bibfnamefont {P.~A.}\ \bibnamefont
  {Lee}}, \bibinfo {author} {\bibfnamefont {N.}~\bibnamefont {Nagaosa}},
  \bibinfo {author} {\bibfnamefont {T.-K.}\ \bibnamefont {Ng}}, \ and\ \bibinfo
  {author} {\bibfnamefont {X.-G.}\ \bibnamefont {Wen}},\ }\href {\doibase
  10.1103/PhysRevB.57.6003} {\bibfield  {journal} {\bibinfo  {journal} {Phys.
  Rev. B}\ }\textbf {\bibinfo {volume} {57}},\ \bibinfo {pages} {6003}
  (\bibinfo {year} {1998})}\BibitemShut {NoStop}%
\bibitem [{\citenamefont {Florens}\ and\ \citenamefont
  {Georges}(2002)}]{Florens2002}%
  \BibitemOpen
  \bibfield  {author} {\bibinfo {author} {\bibfnamefont {S.}~\bibnamefont
  {Florens}}\ and\ \bibinfo {author} {\bibfnamefont {A.}~\bibnamefont
  {Georges}},\ }\href {\doibase 10.1103/PhysRevB.66.165111} {\bibfield
  {journal} {\bibinfo  {journal} {Phys. Rev. B}\ }\textbf {\bibinfo {volume}
  {66}},\ \bibinfo {pages} {165111} (\bibinfo {year} {2002})}\BibitemShut
  {NoStop}%
\bibitem [{\citenamefont {Florens}\ and\ \citenamefont
  {Georges}(2004)}]{Florens2004}%
  \BibitemOpen
  \bibfield  {author} {\bibinfo {author} {\bibfnamefont {S.}~\bibnamefont
  {Florens}}\ and\ \bibinfo {author} {\bibfnamefont {A.}~\bibnamefont
  {Georges}},\ }\href {\doibase 10.1103/PhysRevB.70.035114} {\bibfield
  {journal} {\bibinfo  {journal} {Phys. Rev. B}\ }\textbf {\bibinfo {volume}
  {70}},\ \bibinfo {pages} {035114} (\bibinfo {year} {2004})}\BibitemShut
  {NoStop}%
\bibitem [{\citenamefont {Swingle}(2010)}]{Swingle2010}%
  \BibitemOpen
  \bibfield  {author} {\bibinfo {author} {\bibfnamefont {B.}~\bibnamefont
  {Swingle}},\ }\href {\doibase 10.1103/PhysRevLett.105.050502} {\bibfield
  {journal} {\bibinfo  {journal} {Phys. Rev. Lett.}\ }\textbf {\bibinfo
  {volume} {105}},\ \bibinfo {pages} {050502} (\bibinfo {year}
  {2010})}\BibitemShut {NoStop}%
\bibitem [{\citenamefont {Zhang}\ \emph {et~al.}(2011)\citenamefont {Zhang},
  \citenamefont {Grover},\ and\ \citenamefont {Vishwanath}}]{Zhang2011}%
  \BibitemOpen
  \bibfield  {author} {\bibinfo {author} {\bibfnamefont {Y.}~\bibnamefont
  {Zhang}}, \bibinfo {author} {\bibfnamefont {T.}~\bibnamefont {Grover}}, \
  and\ \bibinfo {author} {\bibfnamefont {A.}~\bibnamefont {Vishwanath}},\
  }\href {\doibase 10.1103/PhysRevLett.107.067202} {\bibfield  {journal}
  {\bibinfo  {journal} {Phys. Rev. Lett.}\ }\textbf {\bibinfo {volume} {107}},\
  \bibinfo {pages} {067202} (\bibinfo {year} {2011})}\BibitemShut {NoStop}%
\bibitem [{\citenamefont {Anderson}(1961)}]{Anderson1961}%
  \BibitemOpen
  \bibfield  {author} {\bibinfo {author} {\bibfnamefont {P.~W.}\ \bibnamefont
  {Anderson}},\ }\href {\doibase 10.1103/PhysRev.124.41} {\bibfield  {journal}
  {\bibinfo  {journal} {Phys. Rev.}\ }\textbf {\bibinfo {volume} {124}},\
  \bibinfo {pages} {41} (\bibinfo {year} {1961})}\BibitemShut {NoStop}%
\bibitem [{\citenamefont {Kondo}(1964)}]{Kondo1964}%
  \BibitemOpen
  \bibfield  {author} {\bibinfo {author} {\bibfnamefont {J.}~\bibnamefont
  {Kondo}},\ }\href {\doibase 10.1143/PTP.32.37} {\bibfield  {journal}
  {\bibinfo  {journal} {Progress of Theoretical Physics}\ }\textbf {\bibinfo
  {volume} {32}},\ \bibinfo {pages} {37} (\bibinfo {year} {1964})}\BibitemShut
  {NoStop}%
\bibitem [{\citenamefont {Hewson}(1997)}]{Hewson1997}%
  \BibitemOpen
  \bibfield  {author} {\bibinfo {author} {\bibfnamefont {A.}~\bibnamefont
  {Hewson}},\ }\href {https://books.google.com/books?id=fPzgHneNFDAC} {\emph
  {\bibinfo {title} {The Kondo Problem to Heavy Fermions}}},\ Cambridge Studies
  in Magnetism\ (\bibinfo  {publisher} {Cambridge University Press},\ \bibinfo
  {year} {1997})\BibitemShut {NoStop}%
\bibitem [{\citenamefont {Chen}\ \emph {et~al.}(2022)\citenamefont {Chen},
  \citenamefont {He}, \citenamefont {Ruan}, \citenamefont {Hwang},
  \citenamefont {Tang}, \citenamefont {Lee}, \citenamefont {Wu}, \citenamefont
  {Zhu}, \citenamefont {Zhang}, \citenamefont {Ryu}, \citenamefont {Wang},
  \citenamefont {Louie}, \citenamefont {Shen}, \citenamefont {Mo},
  \citenamefont {Lee},\ and\ \citenamefont {Crommie}}]{Chen2022}%
  \BibitemOpen
  \bibfield  {author} {\bibinfo {author} {\bibfnamefont {Y.}~\bibnamefont
  {Chen}}, \bibinfo {author} {\bibfnamefont {W.-Y.}\ \bibnamefont {He}},
  \bibinfo {author} {\bibfnamefont {W.}~\bibnamefont {Ruan}}, \bibinfo {author}
  {\bibfnamefont {J.}~\bibnamefont {Hwang}}, \bibinfo {author} {\bibfnamefont
  {S.}~\bibnamefont {Tang}}, \bibinfo {author} {\bibfnamefont {R.~L.}\
  \bibnamefont {Lee}}, \bibinfo {author} {\bibfnamefont {M.}~\bibnamefont
  {Wu}}, \bibinfo {author} {\bibfnamefont {T.}~\bibnamefont {Zhu}}, \bibinfo
  {author} {\bibfnamefont {C.}~\bibnamefont {Zhang}}, \bibinfo {author}
  {\bibfnamefont {H.}~\bibnamefont {Ryu}}, \bibinfo {author} {\bibfnamefont
  {F.}~\bibnamefont {Wang}}, \bibinfo {author} {\bibfnamefont {S.~G.}\
  \bibnamefont {Louie}}, \bibinfo {author} {\bibfnamefont {Z.-X.}\ \bibnamefont
  {Shen}}, \bibinfo {author} {\bibfnamefont {S.-K.}\ \bibnamefont {Mo}},
  \bibinfo {author} {\bibfnamefont {P.~A.}\ \bibnamefont {Lee}}, \ and\
  \bibinfo {author} {\bibfnamefont {M.~F.}\ \bibnamefont {Crommie}},\ }\href
  {\doibase 10.1038/s41567-022-01751-4} {\bibfield  {journal} {\bibinfo
  {journal} {Nature Physics}\ }\textbf {\bibinfo {volume} {18}},\ \bibinfo
  {pages} {1335} (\bibinfo {year} {2022})}\BibitemShut {NoStop}%
\bibitem [{\citenamefont {He}\ and\ \citenamefont {Lee}(2022)}]{He2022}%
  \BibitemOpen
  \bibfield  {author} {\bibinfo {author} {\bibfnamefont {W.-Y.}\ \bibnamefont
  {He}}\ and\ \bibinfo {author} {\bibfnamefont {P.~A.}\ \bibnamefont {Lee}},\
  }\href {\doibase 10.1103/PhysRevB.105.195156} {\bibfield  {journal} {\bibinfo
   {journal} {Phys. Rev. B}\ }\textbf {\bibinfo {volume} {105}},\ \bibinfo
  {pages} {195156} (\bibinfo {year} {2022})}\BibitemShut {NoStop}%
\bibitem [{\citenamefont {Coleman}\ \emph {et~al.}(2005)\citenamefont
  {Coleman}, \citenamefont {Marston},\ and\ \citenamefont
  {Schofield}}]{Coleman2005}%
  \BibitemOpen
  \bibfield  {author} {\bibinfo {author} {\bibfnamefont {P.}~\bibnamefont
  {Coleman}}, \bibinfo {author} {\bibfnamefont {J.~B.}\ \bibnamefont
  {Marston}}, \ and\ \bibinfo {author} {\bibfnamefont {A.~J.}\ \bibnamefont
  {Schofield}},\ }\href {\doibase 10.1103/PhysRevB.72.245111} {\bibfield
  {journal} {\bibinfo  {journal} {Phys. Rev. B}\ }\textbf {\bibinfo {volume}
  {72}},\ \bibinfo {pages} {245111} (\bibinfo {year} {2005})}\BibitemShut
  {NoStop}%
\bibitem [{\citenamefont {Coleman}(1983)}]{Coleman1983}%
  \BibitemOpen
  \bibfield  {author} {\bibinfo {author} {\bibfnamefont {P.}~\bibnamefont
  {Coleman}},\ }\href {\doibase 10.1103/PhysRevB.28.5255} {\bibfield  {journal}
  {\bibinfo  {journal} {Phys. Rev. B}\ }\textbf {\bibinfo {volume} {28}},\
  \bibinfo {pages} {5255} (\bibinfo {year} {1983})}\BibitemShut {NoStop}%
\bibitem [{\citenamefont {Coleman}(2015)}]{Coleman2015}%
  \BibitemOpen
  \bibfield  {author} {\bibinfo {author} {\bibfnamefont {P.}~\bibnamefont
  {Coleman}},\ }\href@noop {} {\emph {\bibinfo {title} {Introduction to
  Many-Body Physics}}}\ (\bibinfo  {publisher} {Cambridge University Press},\
  \bibinfo {year} {2015})\BibitemShut {NoStop}%
\bibitem [{\citenamefont {Burdin}\ \emph {et~al.}(2002)\citenamefont {Burdin},
  \citenamefont {Grempel},\ and\ \citenamefont {Georges}}]{Burdin2002}%
  \BibitemOpen
  \bibfield  {author} {\bibinfo {author} {\bibfnamefont {S.}~\bibnamefont
  {Burdin}}, \bibinfo {author} {\bibfnamefont {D.~R.}\ \bibnamefont {Grempel}},
  \ and\ \bibinfo {author} {\bibfnamefont {A.}~\bibnamefont {Georges}},\ }\href
  {\doibase 10.1103/PhysRevB.66.045111} {\bibfield  {journal} {\bibinfo
  {journal} {Phys. Rev. B}\ }\textbf {\bibinfo {volume} {66}},\ \bibinfo
  {pages} {045111} (\bibinfo {year} {2002})}\BibitemShut {NoStop}%
\bibitem [{\citenamefont {Senthil}\ \emph {et~al.}(2004)\citenamefont
  {Senthil}, \citenamefont {Vojta},\ and\ \citenamefont
  {Sachdev}}]{Senthil2004}%
  \BibitemOpen
  \bibfield  {author} {\bibinfo {author} {\bibfnamefont {T.}~\bibnamefont
  {Senthil}}, \bibinfo {author} {\bibfnamefont {M.}~\bibnamefont {Vojta}}, \
  and\ \bibinfo {author} {\bibfnamefont {S.}~\bibnamefont {Sachdev}},\ }\href
  {\doibase 10.1103/PhysRevB.69.035111} {\bibfield  {journal} {\bibinfo
  {journal} {Phys. Rev. B}\ }\textbf {\bibinfo {volume} {69}},\ \bibinfo
  {pages} {035111} (\bibinfo {year} {2004})}\BibitemShut {NoStop}%
\bibitem [{\citenamefont {Sachdev}(2023)}]{Sachdev2023}%
  \BibitemOpen
  \bibfield  {author} {\bibinfo {author} {\bibfnamefont {S.}~\bibnamefont
  {Sachdev}},\ }\href@noop {} {\emph {\bibinfo {title} {Quantum Phases of
  Matter}}}\ (\bibinfo  {publisher} {Cambridge University Press},\ \bibinfo
  {year} {2023})\BibitemShut {NoStop}%
\bibitem [{\citenamefont {Senthil}(2008)}]{Senthil2008}%
  \BibitemOpen
  \bibfield  {author} {\bibinfo {author} {\bibfnamefont {T.}~\bibnamefont
  {Senthil}},\ }\href {\doibase 10.1103/PhysRevB.78.045109} {\bibfield
  {journal} {\bibinfo  {journal} {Phys. Rev. B}\ }\textbf {\bibinfo {volume}
  {78}},\ \bibinfo {pages} {045109} (\bibinfo {year} {2008})}\BibitemShut
  {NoStop}%
\bibitem [{\citenamefont {Podolsky}\ \emph {et~al.}(2009)\citenamefont
  {Podolsky}, \citenamefont {Paramekanti}, \citenamefont {Kim},\ and\
  \citenamefont {Senthil}}]{Podolsky2009}%
  \BibitemOpen
  \bibfield  {author} {\bibinfo {author} {\bibfnamefont {D.}~\bibnamefont
  {Podolsky}}, \bibinfo {author} {\bibfnamefont {A.}~\bibnamefont
  {Paramekanti}}, \bibinfo {author} {\bibfnamefont {Y.~B.}\ \bibnamefont
  {Kim}}, \ and\ \bibinfo {author} {\bibfnamefont {T.}~\bibnamefont
  {Senthil}},\ }\href {\doibase 10.1103/PhysRevLett.102.186401} {\bibfield
  {journal} {\bibinfo  {journal} {Phys. Rev. Lett.}\ }\textbf {\bibinfo
  {volume} {102}},\ \bibinfo {pages} {186401} (\bibinfo {year}
  {2009})}\BibitemShut {NoStop}%
\bibitem [{\citenamefont {Hermele}\ \emph {et~al.}(2004)\citenamefont
  {Hermele}, \citenamefont {Senthil}, \citenamefont {Fisher}, \citenamefont
  {Lee}, \citenamefont {Nagaosa},\ and\ \citenamefont {Wen}}]{Hermele2004}%
  \BibitemOpen
  \bibfield  {author} {\bibinfo {author} {\bibfnamefont {M.}~\bibnamefont
  {Hermele}}, \bibinfo {author} {\bibfnamefont {T.}~\bibnamefont {Senthil}},
  \bibinfo {author} {\bibfnamefont {M.~P.~A.}\ \bibnamefont {Fisher}}, \bibinfo
  {author} {\bibfnamefont {P.~A.}\ \bibnamefont {Lee}}, \bibinfo {author}
  {\bibfnamefont {N.}~\bibnamefont {Nagaosa}}, \ and\ \bibinfo {author}
  {\bibfnamefont {X.-G.}\ \bibnamefont {Wen}},\ }\href {\doibase
  10.1103/PhysRevB.70.214437} {\bibfield  {journal} {\bibinfo  {journal} {Phys.
  Rev. B}\ }\textbf {\bibinfo {volume} {70}},\ \bibinfo {pages} {214437}
  (\bibinfo {year} {2004})}\BibitemShut {NoStop}%
\bibitem [{\citenamefont {Lee}(2009)}]{Lee2009}%
  \BibitemOpen
  \bibfield  {author} {\bibinfo {author} {\bibfnamefont {S.-S.}\ \bibnamefont
  {Lee}},\ }\href {\doibase 10.1103/PhysRevB.80.165102} {\bibfield  {journal}
  {\bibinfo  {journal} {Phys. Rev. B}\ }\textbf {\bibinfo {volume} {80}},\
  \bibinfo {pages} {165102} (\bibinfo {year} {2009})}\BibitemShut {NoStop}%
\bibitem [{\citenamefont {Salpeter}\ and\ \citenamefont
  {Bethe}(1951)}]{Salpeter1951}%
  \BibitemOpen
  \bibfield  {author} {\bibinfo {author} {\bibfnamefont {E.~E.}\ \bibnamefont
  {Salpeter}}\ and\ \bibinfo {author} {\bibfnamefont {H.~A.}\ \bibnamefont
  {Bethe}},\ }\href {\doibase 10.1103/PhysRev.84.1232} {\bibfield  {journal}
  {\bibinfo  {journal} {Phys. Rev.}\ }\textbf {\bibinfo {volume} {84}},\
  \bibinfo {pages} {1232} (\bibinfo {year} {1951})}\BibitemShut {NoStop}%
\bibitem [{\citenamefont {Greiner}\ and\ \citenamefont
  {Reinhardt}(2003)}]{Greiner2003}%
  \BibitemOpen
  \bibfield  {author} {\bibinfo {author} {\bibfnamefont {W.}~\bibnamefont
  {Greiner}}\ and\ \bibinfo {author} {\bibfnamefont {J.}~\bibnamefont
  {Reinhardt}},\ }\href {https://books.google.de/books?id=Ci-9XMwzkmoC} {\emph
  {\bibinfo {title} {Quantum Electrodynamics}}},\ Physics and astronomy online
  library\ (\bibinfo  {publisher} {Springer},\ \bibinfo {year}
  {2003})\BibitemShut {NoStop}%
\bibitem [{\citenamefont {Lurie}(1968)}]{Lurie1968}%
  \BibitemOpen
  \bibfield  {author} {\bibinfo {author} {\bibfnamefont {D.}~\bibnamefont
  {Lurie}},\ }\href {https://books.google.de/books?id=7NpzjwEACAAJ} {\emph
  {\bibinfo {title} {Particles and Fields}}}\ (\bibinfo  {publisher}
  {Interscience Publishers},\ \bibinfo {year} {1968})\BibitemShut {NoStop}%
\bibitem [{\citenamefont {Banerjee}\ \emph {et~al.}(2016)\citenamefont
  {Banerjee}, \citenamefont {Bridges}, \citenamefont {Yan}, \citenamefont
  {Aczel}, \citenamefont {Li}, \citenamefont {Stone}, \citenamefont {Granroth},
  \citenamefont {Lumsden}, \citenamefont {Yiu}, \citenamefont {Knolle},
  \citenamefont {Bhattacharjee}, \citenamefont {Kovrizhin}, \citenamefont
  {Moessner}, \citenamefont {Tennant}, \citenamefont {Mandrus},\ and\
  \citenamefont {Nagler}}]{Banerjee2016}%
  \BibitemOpen
  \bibfield  {author} {\bibinfo {author} {\bibfnamefont {A.}~\bibnamefont
  {Banerjee}}, \bibinfo {author} {\bibfnamefont {C.~A.}\ \bibnamefont
  {Bridges}}, \bibinfo {author} {\bibfnamefont {J.-Q.}\ \bibnamefont {Yan}},
  \bibinfo {author} {\bibfnamefont {A.~A.}\ \bibnamefont {Aczel}}, \bibinfo
  {author} {\bibfnamefont {L.}~\bibnamefont {Li}}, \bibinfo {author}
  {\bibfnamefont {M.~B.}\ \bibnamefont {Stone}}, \bibinfo {author}
  {\bibfnamefont {G.~E.}\ \bibnamefont {Granroth}}, \bibinfo {author}
  {\bibfnamefont {M.~D.}\ \bibnamefont {Lumsden}}, \bibinfo {author}
  {\bibfnamefont {Y.}~\bibnamefont {Yiu}}, \bibinfo {author} {\bibfnamefont
  {J.}~\bibnamefont {Knolle}}, \bibinfo {author} {\bibfnamefont
  {S.}~\bibnamefont {Bhattacharjee}}, \bibinfo {author} {\bibfnamefont {D.~L.}\
  \bibnamefont {Kovrizhin}}, \bibinfo {author} {\bibfnamefont {R.}~\bibnamefont
  {Moessner}}, \bibinfo {author} {\bibfnamefont {D.~A.}\ \bibnamefont
  {Tennant}}, \bibinfo {author} {\bibfnamefont {D.~G.}\ \bibnamefont
  {Mandrus}}, \ and\ \bibinfo {author} {\bibfnamefont {S.~E.}\ \bibnamefont
  {Nagler}},\ }\href {\doibase 10.1038/nmat4604} {\bibfield  {journal}
  {\bibinfo  {journal} {Nature Materials}\ }\textbf {\bibinfo {volume} {15}},\
  \bibinfo {pages} {733} (\bibinfo {year} {2016})}\BibitemShut {NoStop}%
\bibitem [{\citenamefont {Ribak}\ \emph {et~al.}(2017)\citenamefont {Ribak},
  \citenamefont {Silber}, \citenamefont {Baines}, \citenamefont {Chashka},
  \citenamefont {Salman}, \citenamefont {Dagan},\ and\ \citenamefont
  {Kanigel}}]{Ribak2017}%
  \BibitemOpen
  \bibfield  {author} {\bibinfo {author} {\bibfnamefont {A.}~\bibnamefont
  {Ribak}}, \bibinfo {author} {\bibfnamefont {I.}~\bibnamefont {Silber}},
  \bibinfo {author} {\bibfnamefont {C.}~\bibnamefont {Baines}}, \bibinfo
  {author} {\bibfnamefont {K.}~\bibnamefont {Chashka}}, \bibinfo {author}
  {\bibfnamefont {Z.}~\bibnamefont {Salman}}, \bibinfo {author} {\bibfnamefont
  {Y.}~\bibnamefont {Dagan}}, \ and\ \bibinfo {author} {\bibfnamefont
  {A.}~\bibnamefont {Kanigel}},\ }\href {\doibase 10.1103/PhysRevB.96.195131}
  {\bibfield  {journal} {\bibinfo  {journal} {Phys. Rev. B}\ }\textbf {\bibinfo
  {volume} {96}},\ \bibinfo {pages} {195131} (\bibinfo {year}
  {2017})}\BibitemShut {NoStop}%
\bibitem [{\citenamefont {Murayama}\ \emph {et~al.}(2020)\citenamefont
  {Murayama}, \citenamefont {Sato}, \citenamefont {Taniguchi}, \citenamefont
  {Kurihara}, \citenamefont {Xing}, \citenamefont {Huang}, \citenamefont
  {Kasahara}, \citenamefont {Kasahara}, \citenamefont {Kimchi}, \citenamefont
  {Yoshida}, \citenamefont {Iwasa}, \citenamefont {Mizukami}, \citenamefont
  {Shibauchi}, \citenamefont {Konczykowski},\ and\ \citenamefont
  {Matsuda}}]{Murayama2020}%
  \BibitemOpen
  \bibfield  {author} {\bibinfo {author} {\bibfnamefont {H.}~\bibnamefont
  {Murayama}}, \bibinfo {author} {\bibfnamefont {Y.}~\bibnamefont {Sato}},
  \bibinfo {author} {\bibfnamefont {T.}~\bibnamefont {Taniguchi}}, \bibinfo
  {author} {\bibfnamefont {R.}~\bibnamefont {Kurihara}}, \bibinfo {author}
  {\bibfnamefont {X.~Z.}\ \bibnamefont {Xing}}, \bibinfo {author}
  {\bibfnamefont {W.}~\bibnamefont {Huang}}, \bibinfo {author} {\bibfnamefont
  {S.}~\bibnamefont {Kasahara}}, \bibinfo {author} {\bibfnamefont
  {Y.}~\bibnamefont {Kasahara}}, \bibinfo {author} {\bibfnamefont
  {I.}~\bibnamefont {Kimchi}}, \bibinfo {author} {\bibfnamefont
  {M.}~\bibnamefont {Yoshida}}, \bibinfo {author} {\bibfnamefont
  {Y.}~\bibnamefont {Iwasa}}, \bibinfo {author} {\bibfnamefont
  {Y.}~\bibnamefont {Mizukami}}, \bibinfo {author} {\bibfnamefont
  {T.}~\bibnamefont {Shibauchi}}, \bibinfo {author} {\bibfnamefont
  {M.}~\bibnamefont {Konczykowski}}, \ and\ \bibinfo {author} {\bibfnamefont
  {Y.}~\bibnamefont {Matsuda}},\ }\href {\doibase
  10.1103/PhysRevResearch.2.013099} {\bibfield  {journal} {\bibinfo  {journal}
  {Phys. Rev. Res.}\ }\textbf {\bibinfo {volume} {2}},\ \bibinfo {pages}
  {013099} (\bibinfo {year} {2020})}\BibitemShut {NoStop}%
\bibitem [{\citenamefont {Hermele}(2007)}]{Hermele2007}%
  \BibitemOpen
  \bibfield  {author} {\bibinfo {author} {\bibfnamefont {M.}~\bibnamefont
  {Hermele}},\ }\href {\doibase 10.1103/PhysRevB.76.035125} {\bibfield
  {journal} {\bibinfo  {journal} {Phys. Rev. B}\ }\textbf {\bibinfo {volume}
  {76}},\ \bibinfo {pages} {035125} (\bibinfo {year} {2007})}\BibitemShut
  {NoStop}%
\bibitem [{\citenamefont {Chen}\ \emph {et~al.}(2012)\citenamefont {Chen},
  \citenamefont {Essin},\ and\ \citenamefont {Hermele}}]{Chen2012}%
  \BibitemOpen
  \bibfield  {author} {\bibinfo {author} {\bibfnamefont {G.}~\bibnamefont
  {Chen}}, \bibinfo {author} {\bibfnamefont {A.}~\bibnamefont {Essin}}, \ and\
  \bibinfo {author} {\bibfnamefont {M.}~\bibnamefont {Hermele}},\ }\href
  {\doibase 10.1103/PhysRevB.85.094418} {\bibfield  {journal} {\bibinfo
  {journal} {Phys. Rev. B}\ }\textbf {\bibinfo {volume} {85}},\ \bibinfo
  {pages} {094418} (\bibinfo {year} {2012})}\BibitemShut {NoStop}%
\bibitem [{\citenamefont {Kitaev}(2003)}]{Kitaev2003}%
  \BibitemOpen
  \bibfield  {author} {\bibinfo {author} {\bibfnamefont {A.}~\bibnamefont
  {Kitaev}},\ }\href {\doibase https://doi.org/10.1016/S0003-4916(02)00018-0}
  {\bibfield  {journal} {\bibinfo  {journal} {Annals of Physics}\ }\textbf
  {\bibinfo {volume} {303}},\ \bibinfo {pages} {2} (\bibinfo {year}
  {2003})}\BibitemShut {NoStop}%
\bibitem [{\citenamefont {Nayak}\ \emph {et~al.}(2008)\citenamefont {Nayak},
  \citenamefont {Simon}, \citenamefont {Stern}, \citenamefont {Freedman},\ and\
  \citenamefont {Das~Sarma}}]{Nayak2008}%
  \BibitemOpen
  \bibfield  {author} {\bibinfo {author} {\bibfnamefont {C.}~\bibnamefont
  {Nayak}}, \bibinfo {author} {\bibfnamefont {S.~H.}\ \bibnamefont {Simon}},
  \bibinfo {author} {\bibfnamefont {A.}~\bibnamefont {Stern}}, \bibinfo
  {author} {\bibfnamefont {M.}~\bibnamefont {Freedman}}, \ and\ \bibinfo
  {author} {\bibfnamefont {S.}~\bibnamefont {Das~Sarma}},\ }\href {\doibase
  10.1103/RevModPhys.80.1083} {\bibfield  {journal} {\bibinfo  {journal} {Rev.
  Mod. Phys.}\ }\textbf {\bibinfo {volume} {80}},\ \bibinfo {pages} {1083}
  (\bibinfo {year} {2008})}\BibitemShut {NoStop}%
\bibitem [{\citenamefont {Senthil}\ and\ \citenamefont
  {Fisher}(2000)}]{Senthil2000}%
  \BibitemOpen
  \bibfield  {author} {\bibinfo {author} {\bibfnamefont {T.}~\bibnamefont
  {Senthil}}\ and\ \bibinfo {author} {\bibfnamefont {M.~P.~A.}\ \bibnamefont
  {Fisher}},\ }\href {\doibase 10.1103/PhysRevB.62.7850} {\bibfield  {journal}
  {\bibinfo  {journal} {Phys. Rev. B}\ }\textbf {\bibinfo {volume} {62}},\
  \bibinfo {pages} {7850} (\bibinfo {year} {2000})}\BibitemShut {NoStop}%
\bibitem [{\citenamefont {Qi}\ and\ \citenamefont {Sachdev}(2010)}]{Qi2010}%
  \BibitemOpen
  \bibfield  {author} {\bibinfo {author} {\bibfnamefont {Y.}~\bibnamefont
  {Qi}}\ and\ \bibinfo {author} {\bibfnamefont {S.}~\bibnamefont {Sachdev}},\
  }\href {\doibase 10.1103/PhysRevB.81.115129} {\bibfield  {journal} {\bibinfo
  {journal} {Phys. Rev. B}\ }\textbf {\bibinfo {volume} {81}},\ \bibinfo
  {pages} {115129} (\bibinfo {year} {2010})}\BibitemShut {NoStop}%
\bibitem [{\citenamefont {Sachdev}(2018)}]{Sachdev2018}%
  \BibitemOpen
  \bibfield  {author} {\bibinfo {author} {\bibfnamefont {S.}~\bibnamefont
  {Sachdev}},\ }\href {\doibase 10.1088/1361-6633/aae110} {\bibfield  {journal}
  {\bibinfo  {journal} {Reports on Progress in Physics}\ }\textbf {\bibinfo
  {volume} {82}},\ \bibinfo {pages} {014001} (\bibinfo {year}
  {2018})}\BibitemShut {NoStop}%
\end{thebibliography}
%

\cleardoublepage{}

\onecolumngrid

\appendix

\section{Derivation of the slave rotor Hamiltonian in canonical form}

According to \citep{Florens2002,Florens2004}, the Hamiltonian of
the U(1) quantum spin liquid with Anderson impurity lattice can be
represented using the slave rotor formalism. The Hamiltonian is given
by:

\begin{align}
H= & \sum_{ij\sigma}t_{ij}c_{i,\sigma}^{\dagger}c_{j,\sigma}+\sum_{i\sigma}\epsilon_{d}d_{i,\sigma}^{\dagger}d_{i,\sigma}+V\sum_{i\sigma}c_{i,\sigma}^{\dagger}d_{i,\sigma}+h.c.+\frac{U_{QSL}}{2}\sum_{i}\left(\sum_{\sigma}c_{i,\sigma}^{\dagger}c_{i,\sigma}-1\right)^{2}+\frac{U}{2}\sum_{i}\left(\sum_{\sigma}d_{i,\sigma}^{\dagger}d_{i,\sigma}-1\right)^{2}\nonumber \\
= & \sum_{ij\sigma}t_{ij}f_{i,\sigma}^{\dagger}f_{j,\sigma}X_{j}^{\dagger}X_{i}+h.c.-\sum_{i\sigma}\left(\mu_{0}+h_{1,i}\right)f_{i,\sigma}^{\dagger}f_{i,\sigma}+\sum_{i\sigma}\left(\epsilon_{0}-h_{2}\right)a_{i,\sigma}^{\dagger}a_{i,\sigma}+V\sum_{i\sigma}f_{i,\sigma}^{\dagger}a_{i,\sigma}Y_{i}^{\dagger}X_{i}+h.c.\nonumber \\
 & +\frac{U_{QSL}}{4}\sum_{i}L_{X,i}^{2}+\frac{U}{4}\sum_{i}L_{Y,i}^{2}.\label{Original slave rotor Hamiltonian}
\end{align}

To compute the angular momentum in the Hamiltonian, it is useful to
first consider a two-dimensional rotor described by the coordinates
$\left(x_{i},y_{i}\right)$ with the constraint $x_{i}^{2}+y_{i}^{2}=1$.
The O(2) rotor is equivalent to the U(1) rotor, allowing the real
space coordinates to be represented by complex numbers $X_{i}=x_{i}+iy_{i}$.
The relationships between the coordinate operators, derivative operators,
and momentum operators in real and complex coordinates are as follows:

\begin{align}
\left(\begin{array}{c}
x_{i}\\
y_{i}
\end{array}\right)=\left(\begin{array}{cc}
\frac{1}{2} & \frac{1}{2}\\
\frac{1}{2i} & -\frac{1}{2i}
\end{array}\right)\left(\begin{array}{c}
X_{i}\\
X_{i}^{\dagger}
\end{array}\right), & \frac{\partial}{\partial\left(\begin{array}{c}
X_{i}\\
X_{i}^{\dagger}
\end{array}\right)}=\left(\begin{array}{cc}
\frac{1}{2} & \frac{1}{2i}\\
\frac{1}{2} & -\frac{1}{2i}
\end{array}\right)\frac{\partial}{\partial\left(\begin{array}{c}
x_{i}\\
y_{i}
\end{array}\right)},\nonumber \\
\frac{\partial}{\partial\left(\begin{array}{c}
x_{i}\\
y_{i}
\end{array}\right)}=\left(\begin{array}{cc}
1 & 1\\
i & -i
\end{array}\right)\frac{\partial}{\partial\left(\begin{array}{c}
X_{i}\\
X_{i}^{\dagger}
\end{array}\right)}, & \left(\begin{array}{c}
p_{x,i}\\
p_{y,i}
\end{array}\right)=\left(\begin{array}{cc}
1 & 1\\
i & -i
\end{array}\right)\left(\begin{array}{c}
P_{i}\\
P_{i}^{\dagger}
\end{array}\right).
\end{align}

Additionally, the angular momentum operator in complex coordinate
is: 
\begin{equation}
L_{X,i}=x_{i}p_{y,i}-p_{x,i}y_{i}=i\left(X_{i}P_{i}-X_{i}^{\dagger}P_{i}^{\dagger}\right),
\end{equation}

\noindent and the square of the angular momentum operator can be expressed
in terms of the momentum operators:

\noindent 
\begin{equation}
L_{X,i}^{2}\rightarrow p_{x,i}^{2}+p_{y,i}^{2}=\left(P_{i}+P_{i}^{\dagger}\right)^{2}-\left(P_{i}-P_{i}^{\dagger}\right)^{2}=4P_{i}P_{i}^{\dagger}.
\end{equation}

By incorporating these expressions similar ones for operator $L_{Y,i}$
into the original slave rotor Hamiltonian Eq. \eqref{Original slave rotor Hamiltonian},
we arrive at the final form of the slave rotor Hamiltonian Eq. \eqref{final slave rotor H}.
It is important to note that the coefficients $\frac{U}{2}$ and $\frac{U_{QSL}}{2}$
have been redefined as $\frac{U}{4}$ and $\frac{U_{QSL}}{4}$, respectively,
to ensure the correct atomic limit, as pointed out in \citep{Florens2002}.

\section{Derivation of the self-consistent equations\label{Appendix B}}

\phantom{}

To calculate the Green's functions for spinons and chargons within
the Hamiltonian denoted by Eqs. \eqref{eq:QSL slave rotor H}-\eqref{Hc},
we consider their respective equations of motion separately.

For the spinon Green's function, the equation of motion is given
by:

\begin{equation}
\left(\begin{array}{cc}
i\omega_{n}-\epsilon_{0} & -w\\
-w & i\omega_{n}-h_{f}(\boldsymbol{k})
\end{array}\right)\left(\begin{array}{c}
G(a,a^{\dagger},\sigma,i\omega_{n})\\
G(f,a^{\dagger},i\omega_{n},\boldsymbol{k},\sigma)
\end{array}\right)=\left(\begin{array}{c}
1\\
0
\end{array}\right).
\end{equation}

Solving these linear equations yields the Green's functions for the
spinon: Eqs. \eqref{GFff0}-\eqref{GFaf}:

\begin{equation}
G(a,a^{\dagger},i\omega_{n},\sigma)=\frac{1}{i\omega_{n}-\epsilon_{d}-\frac{w^{2}}{i\omega_{n}-h_{f}(\boldsymbol{k})}},
\end{equation}

\noindent 
\begin{equation}
G(f,a^{\dagger},i\omega_{n},\boldsymbol{k},\sigma)=\frac{wVG^{0}(f,f^{\dagger},i\omega_{n},\boldsymbol{k},\sigma)}{i\omega_{n}-\epsilon_{0}-w^{2}G^{0}(f,f^{\dagger},i\omega_{n},\boldsymbol{k},\sigma)}.\label{GFaf2}
\end{equation}

Similarly, For the chargon Green's function, the equation of motion
is:

\begin{equation}
\ensuremath{\ensuremath{\mathcal{G}^{-1}\left(i\nu_{n},\boldsymbol{k}\right)\left(\begin{array}{c}
G\left(Y,Y^{\dagger},i\nu_{n},\boldsymbol{k}\right)\\
G\left(Q^{\dagger},Y^{\dagger},i\nu_{n},\boldsymbol{k}\right)\\
G\left(X,Y^{\dagger},i\nu_{n},\boldsymbol{k}\right)\\
G\left(P^{\dagger},Y^{\dagger},i\nu_{n},\boldsymbol{k}\right)
\end{array}\right)=\left(\begin{array}{c}
0\\
-i\\
0\\
0
\end{array}\right)}},
\end{equation}

\noindent where $\mathcal{G}^{-1}(i\nu_{n},\boldsymbol{k})$ is the
inverse Green function of the chargon, expressed as:

\noindent 
\begin{equation}
\mathcal{G}^{-1}\left(i\nu_{n},\boldsymbol{k}\right)=i\nu_{n}\mathbf{1}_{4}-i\left(\begin{array}{cccc}
0 & U & 0 & 0\\
-\lambda_{2} & 0 & u & 0\\
0 & 0 & 0 & U_{QSL}\\
u & 0 & -\omega_{X}^{2}(\boldsymbol{k}) & 0
\end{array}\right).
\end{equation}

Following this method, one can derive the Green's function expressions
for the chargon, Eqs. \eqref{GFXX0}-\eqref{GFYY}:

\noindent 
\begin{equation}
G(Y,Y^{\dagger},i\nu_{n},\boldsymbol{k})=\frac{-1}{\frac{\nu_{n}^{2}}{U}+\lambda_{2}+u^{2}G^{0}(X,X^{\dagger},i\nu_{n},\boldsymbol{k})},\label{GFYY2}
\end{equation}

\noindent 
\begin{equation}
G(X,Y^{\dagger},i\nu_{n},\boldsymbol{k})=\frac{uG^{0}(X,X^{\dagger},i\omega_{n},k)}{\frac{\nu_{n}^{2}}{U}+\lambda_{2}+u^{2}G^{0}(X,X^{\dagger},i\nu_{n},\boldsymbol{k})},\label{GFXY2}
\end{equation}

\begin{equation}
G(X,X^{\dagger},i\nu_{n},\boldsymbol{k})=\frac{-1}{\frac{\nu_{n}^{2}}{U_{QSL}}+\omega_{X}^{2}(\boldsymbol{k})+u^{2}G^{0}(Y,Y^{\dagger},i\nu_{n},\boldsymbol{k})}.
\end{equation}

\noindent According to the definitions of $u$ and $w$, and $\left\langle Y^{\dagger}Y\right\rangle =1$,
we can obtain the Eqs. \eqref{SCE u}-\eqref{SCE lambda} in the main
text.

\section{Translate chargon into canonical boson representation}

Plug the expressions for $X_{k}=\frac{\sqrt{U_{QSL}}}{\sqrt{2\epsilon(k)}}\left(h_{k}+d_{k}^{\dagger}\right)$,
$P_{k}=i\sqrt{\frac{\epsilon(k)}{2U_{QSL}}}\left(h_{k}^{\dagger}-d_{k}\right)$
into chargon part Hamiltonian in spin liquid:

\begin{align}
H_{C}= & \sum_{k}\omega^2_{X}(k)X_{k}^{\dagger}X_{k}+U_{QSL}\sum_{k}P_{k}^{\dagger}P_{k}\nonumber \\
= & \sum_{k}\frac{U_{QSL}\omega^2_{X}(k)}{2\epsilon(k)}\left(h_{k}^{\dagger}+d_{k}\right)\left(h_{k}+d_{k}^{\dagger}\right)+\sum_{k}\frac{\epsilon(k)}{2}\left(h_{k}-d_{k}^{\dagger}\right)\left(h_{k}^{\dagger}-d_{k}\right)\nonumber \\
= & \sum_{k}\frac{\epsilon(k)}{2}\left(h_{k}^{\dagger}h_{k}+h_{k}^{\dagger}d_{k}^{\dagger}+d_{k}h_{k}+d_{k}d_{k}^{\dagger}\right)+\sum_{k}\frac{\epsilon(k)}{2}\left(h_{k}h_{k}^{\dagger}-h_{k}d_{k}-d_{k}^{\dagger}h_{k}^{\dagger}+d_{k}^{\dagger}d_{k}\right)\nonumber \\
= & \sum_{k}\epsilon(k)\left(h_{k}^{\dagger}h_{k}+d_{k}^{\dagger}d_{k}+1\right).
\end{align}

\noindent In this equation, $\epsilon(k)=\sqrt{U_{QSL}}\omega_{X}(k)$
represents the chargon dispersion relation, and operators $h_{k}^{(\dagger)}$
and $d_{k}^{(\dagger)}$ are the annihilation (creation) operators
for holons and doublons, respectively.

\section{Emergent U(1) gauge field in quantum spin liquid\label{sec:Emergent--gaugeAppendix D: Emergent U(1) gauge field in quantum spin liquid}}

\phantom{}

In the context of quantum spin liquids, the slave rotor $X_{i}$ and
its associated angular momentum $L_{i}$ can be interpreted as gauge
fields, as discussed in seminal works by \citep{Senthil2000,Qi2010,Sachdev2018}.
When the high-energy degrees of freedom for spinons and chargons are
integrated out at the Gaussian level, a dynamic term for an emergent
U(1) gauge field can be derived:

\begin{align}
S_{G}= & \frac{1}{g^{2}}\int_{0}^{\beta}d\tau\sum_{i}\mathrm{Re}\left[\prod_{\triangle}\exp\left(-iA_{\mu,i}a\right)\right].
\end{align}

\noindent Here, the gauge field $A_{\mu,i}$ arises from the Lagrange
multiplier $h_{i}$ and local gauge redundancy. Furthermore, upon
coarse-graining the lattice model, the continuum low-energy effective
action for the U(1) quantum spin liquid with spinon Fermi surface
is given by \cite{Senthil2008,Podolsky2009,He2022}:

\begin{equation}
S_{\mathrm{QSL}}=S_{S}+S_{C}+S_{M},
\end{equation}

\begin{equation}
S_{S}=\int_{0}^{\beta}d\tau\int d\bm{r}\sum_{\sigma}\left[f_{\sigma,\bm{r}}^{\dagger}\left(\partial_{\tau}-iA_{0}(\boldsymbol{r})-\mu\right)f_{\sigma}(\boldsymbol{r})+\frac{\hbar^{2}}{2m_{f}}\left(\partial_{\bm{r}}+i\bm{A}(\boldsymbol{r})\right)f_{\sigma,\bm{r}}^{\dagger}\cdot\left(\partial_{\bm{r}}-i\bm{A}(\boldsymbol{r})\right)f_{\sigma}(\boldsymbol{r})\right],
\end{equation}

\begin{equation}
S_{C}=\int_{0}^{\beta}d\tau\int d\bm{r}\left[\left(\partial_{\tau}+iA_{0}(\boldsymbol{r})\right)X^{\dagger}(\boldsymbol{r})\left(\partial_{\tau}-iA_{0}(\boldsymbol{r})\right)X(\boldsymbol{r})+\hbar^{2}v_{C}^{2}\left(\partial_{\bm{r}}+i\bm{A}(\boldsymbol{r})\right)X^{\dagger}(\boldsymbol{r})\left(\partial_{\bm{r}}-i\bm{A}(\boldsymbol{r})\right)X(\boldsymbol{r})+\Delta_{g}^{2}X^{\dagger}(\boldsymbol{r})X(\boldsymbol{r})\right],
\end{equation}

\begin{align}
S_{M}= & \frac{1}{2g^{2}}\int_{0}^{\beta}d\tau\int d\bm{r}\left[\left(\nabla A_{0}(\boldsymbol{r})+\partial_{\tau}\bm{A}(\boldsymbol{r})\right)^{2}+\left(\nabla\times\bm{A}(\boldsymbol{r})\right)^{2}\right].
\end{align}

\noindent The coefficients above are as follows: $m_{f}$ represents
the effective mass of the spinon, $v_{C}$ denotes the effective velocity
of the chargon, and $g$ is the effective coupling constant. The magnitude
of g is given by $g^{2}\sim\frac{\Delta_{g}^{2}}{tt_{F}}$ \citep{Lee2005}.

Utilizing the transformation of the chargon field $X_{\bm{r}}$ as
outlined in the previous section Appendix \ref{Appendix B}, we can
derive the action for the chargon in the Schr\"{o}dinger form:

\begin{equation}
S_{C}=\int_{0}^{\beta}\int d\bm{r}\left[h_{\bm{r}}^{\dagger}\left(\partial_{\tau}+iA_{0,\bm{r}}\right)h_{\bm{r}}+d_{\bm{r}}^{\dagger}\left(-\partial_{\tau}-iA_{0,\bm{r}}\right)d_{\bm{r}}+\epsilon(\partial_{\bm{r}}+i\bm{A}_{\bm{r}})h_{\bm{r}}^{\dagger}h_{\bm{r}}+\epsilon(\partial_{\bm{r}}+i\bm{A}_{\bm{r}})d_{\bm{r}}^{\dagger}d_{\bm{r}}\right],
\end{equation}

\noindent where $\epsilon(\boldsymbol{k})=\sqrt{\hbar^{2}v_{C}^{2}\boldsymbol{k}^{2}+\Delta_{g}^{2}}$
is low energy limit of lattice version chargon dispersion.

In the vicinity of the Hubbard band edge energies, the group velocity
exhibits diminutive values, resulting in the transverse component
of gauge field fluctuations arising from the current-current correlation
being negligible. Conversely, the dominant influence stems from the
longitudinal component of gauge field fluctuations. Through the process
of integrating out the gauge field $A_{0,\bm{r}}$, the interaction
term between spinon, chargon and themselves can be derived:

\begin{align}
S_{\textrm{int}}= & \int_{0}^{\beta}d\tau\int d\bm{r}d\boldsymbol{r}^{\prime}V(\boldsymbol{r}-\boldsymbol{r}^{\prime})\sum_{\sigma,\sigma'}\left[f_{\sigma,\bm{r}}^{\dagger}f_{\sigma,\bm{r}}f_{\sigma',\bm{r}'}^{\dagger}f_{\sigma',\bm{r}'}-\left(X_{\bm{r}}P_{\bm{r}}-X_{\bm{r}}^{\dagger}P_{\bm{r}}^{\dagger}\right)\left(X_{\bm{r}'}P_{\bm{r}'}-X_{\bm{r}'}^{\dagger}P_{\bm{r}'}^{\dagger}\right)+2if_{\sigma,\bm{r}}^{\dagger}f_{\sigma,\bm{r}}\left(X_{\bm{r}'}P_{\bm{r}'}-X_{\bm{r}'}^{\dagger}P_{\bm{r}'}^{\dagger}\right)\right]\nonumber \\
= & \int_{0}^{\beta}d\tau\int d\bm{r}d\boldsymbol{r}^{\prime}V(\boldsymbol{r}-\boldsymbol{r}^{\prime})\sum_{\sigma,\sigma'}\left[f_{\sigma,\bm{r}}^{\dagger}f_{\sigma,\bm{r}}f_{\sigma',\bm{r}'}^{\dagger}f_{\sigma',\bm{r}'}+\left(a_{\bm{r}}^{\dagger}a_{\bm{r}}-b_{\bm{r}}^{\dagger}b_{\bm{r}}\right)\left(a_{\bm{r}'}^{\dagger}a_{\bm{r}'}-b_{\bm{r}'}^{\dagger}b_{\bm{r}'}\right)+2f_{\sigma,\bm{r}}^{\dagger}f_{\sigma,\bm{r}}\left(a_{\bm{r}'}^{\dagger}a_{\bm{r}'}-b_{\bm{r}'}^{\dagger}b_{\bm{r}'}\right)\right],
\end{align}

\noindent where Coulomb potential $V(\boldsymbol{r}-\boldsymbol{r}^{\prime})=\frac{g^{2}}{\left|\boldsymbol{r}-\boldsymbol{r}^{\prime}\right|}$.

\section{Calculate the Green's function and density of states of the two-body
bound state by Bethe-Salpeter equation\label{Appendix E:Calculate-the-Green's}}

\phantom{}

In the system considered in this article, the electron at position
$\boldsymbol{r}_{i}$ is decomposed into spinon and chargon at the
same lattice site, so when considering the two particles forming a
bound state, we only need to analyze the coupling at the same location.
Therefore, we start from the Bethe-Salpeter equation satisfied by
two particles with same spatial coordinates:

\begin{align}
G_{2}\left(4;1\right)= & G_{1,0}\left(4,1\right)G_{1,0}\left(4,1\right)+\int dX_{c,2}dX_{c,3}G_{0}\left(4,3\right)G_{0}\left(4,3\right)K^{*}\left(3;2\right)G_{2}\left(2;1\right)\label{eq: X BSE}
\end{align}

\noindent Here, $G_{2}$ is the the two-body Green's function, and
$G_{1,0}$ is the free single-particle Green's function. $X_{c}=\frac{1}{2}(x_{1}+x_{2})$
represents the center coordinate. It is easily known that the system
has a translation invariance, so $G_{2}\left(4;1\right)=G_{2}\left(X_{c,4}-X_{c,1}\right)$,
which is independent of the relative coordinates, $\bar{x}=x_{1}-x_{2}$,
and only depends on the difference of the center coordinates.

Upon transforming to the momentum space, we obtain:

\begin{align}
\delta(P_{c,4}-P_{c,1})G_{2}\left(P_{c,1}\right)= & \delta(P_{c,4}-P_{c,1})G_{0,p}\left(P_{c,1}\right)*G_{0,p}\left(P_{c,1}\right)+[\beta(2\pi)^{2}]^{-2}\int dP_{c,2}dP_{c,3}\delta(P_{c,4}-P_{c,3})G_{0,p}\left(P_{c,3}\right)*G_{0,p}\left(P_{c,3}\right)\nonumber \\
 & \times\delta(P_{c,3}-P_{c,2})K_{p}^{*}\left(P_{c,2}\right)\delta(P_{c,2}-P_{c,1})G_{2}\left(P_{c,1}\right).\label{P BSE}
\end{align}

\noindent In this equation, the integral over the central momentum
$P_{c}$ contains a sum over Matsubara frequencies. The Green's functions
in the expression are independent of the relative momentum $\bar{p}=\frac{1}{2}(p_{1}-p_{2})$,
and it is easy to prove that the momentum of the spinon and chargons
are equal in the system considered in this article.

For instance, in the case of the U(1) spin liquid with spinon Fermi
surface on triangular lattice, we need to replace the Green's functions
in the above equation with those of the free spinons and chargons
Eqs. \eqref{GFff0}-\eqref{GFXX0}, and substitute $G_{2}\left(P_{1}\right)=-G(c,c^{\dagger},i\omega_{n},\boldsymbol{k},\sigma)$.

Simultaneously, when analyzing the bound states of the spinons and
charge carriers, it is necessary to consider the screen from the spinon
Fermi sea. Here, we apply the Lindhard approximation, considering
only the static RPA-corrected emergent Coulomb potential, $V_{s}(\boldsymbol{q})=V_{\mathrm{RPA}}(\omega=0,\boldsymbol{q})=\frac{g^{2}}{\boldsymbol{q}^{2}+\kappa^{2}}$,
where $\kappa=g\sqrt{N_{f}}$ represents the Thomas-Fermi screening,
and $N_{f}=D_{f}(0)$ is the density of states at the Fermi energy
of the spinons.

Additionally, since we only consider interactions between on the same
lattice site, the Yukawa potential could be further simplified to
a local constant interaction \citep{He2022}:

\begin{equation}
V_{\bm{r}}\approx\frac{g^{2}}{\left(2\pi\right)^{2}}\int\frac{1}{g^{2}N_{f}}d\bm{q}=\int\frac{d\bm{q}}{4\pi^{2}N_{f}}\approx\Lambda_{f},
\end{equation}

\noindent where $\varLambda_{f}$ is the spinon half-bandwidth.

Therefore, the two-body kernel from local constant interaction can
be concluded as a constant ladder approximation: $\frac{1}{(2\pi)^{2}}\int dkK^{*}\left(i\nu_{n},\boldsymbol{k}\right)\approx-\frac{i\nu_{n}V_{\boldsymbol{r}}}{U_{QSL}}$
\citep{He2022}. Finally, we obtain the approximate Bethe-Salpeter
equation for host spin liquid:

In the ladder approximation, the Bethe-Salpeter equation is given
by \citep{Salpeter1951,Greiner2003,Lurie1968}:

\begin{align}
G(c,c^{\dagger},i\omega_{n},\boldsymbol{k}_{1},\sigma)= & -\frac{1}{\beta N}\sum_{i\nu_{n},\boldsymbol{k}_{2}}G(f,f^{\dagger},i\omega_{n}+i\nu_{n},\boldsymbol{k}_{1}+\boldsymbol{k}_{2},\sigma)G(X,X^{\dagger},i\nu_{n},\boldsymbol{k}_{1})\nonumber \\
 & -\frac{1}{\beta N}\sum_{i\nu_{n},\boldsymbol{k}_{2}}G(f,f^{\dagger},i\omega_{n}+i\nu_{n},\boldsymbol{k}_{1}+\boldsymbol{k}_{2},\sigma)G(X,X^{\dagger},i\nu_{n},\boldsymbol{k}_{1})\nonumber \\
 & \times i\nu_{n}\frac{V_{\bm{r}}}{U_{QSL}}G(c,c^{\dagger},i\omega_{n},\boldsymbol{k}_{1},\sigma).\label{c BSE GF}
\end{align}

\noindent The simplified corrected electron Green's function is shown
as:

\begin{align}
G(c,c^{\dagger},i\omega_{n},\boldsymbol{k}_{1},\sigma)= & -\frac{1}{\beta N}\sum_{i\nu_{n},\boldsymbol{k}_{1}^{\prime}}G(f,f^{\dagger},i\omega_{n}+i\nu_{n},\boldsymbol{k}_{1}+\boldsymbol{k}_{1}^{\prime},\sigma)G(X,X^{\dagger},i\nu_{n},\boldsymbol{k}_{1}^{\prime})\nonumber \\
 & \times\left[1-\frac{1}{\beta N}\sum_{i\nu_{n},\boldsymbol{k}_{1}^{\prime}}i\nu_{n}\frac{V_{\bm{r}}}{U_{QSL}}G(f,f^{\dagger},i\omega_{n}+i\nu_{n},\boldsymbol{k}_{1}+\boldsymbol{k}_{1}^{\prime},\sigma)G(X,X^{\dagger},i\nu_{n},\boldsymbol{k}_{1}^{\prime})\right]^{-1}\label{c GF}
\end{align}

\begin{figure}
\noindent \begin{centering}
\includegraphics[width=0.6\linewidth]{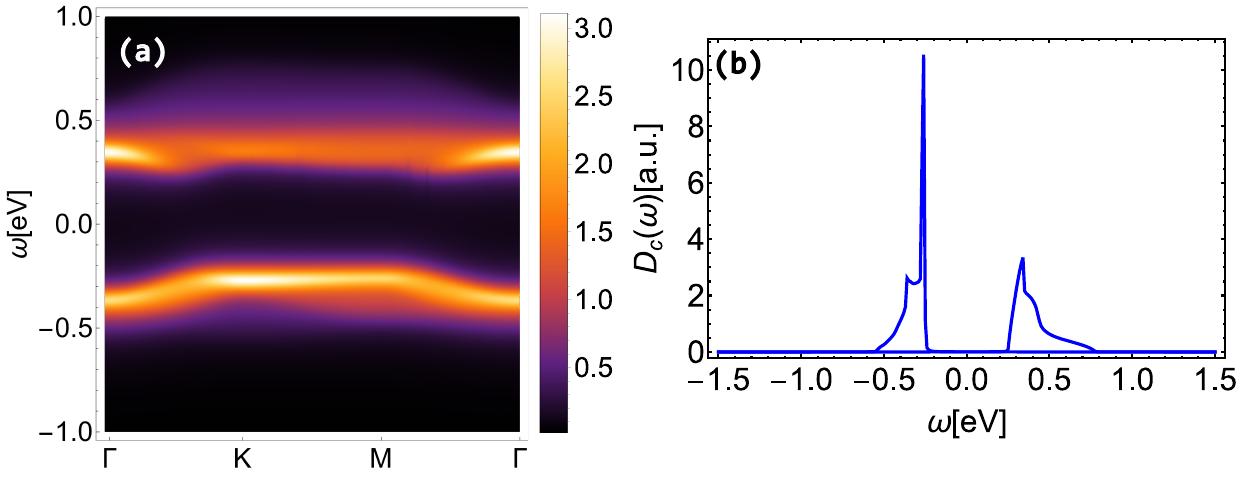} 
\par\end{centering}
\caption{(a) QSL bound state electron spectral density $A_{\mathrm{QSL}}(\omega,\boldsymbol{k})$
along high symmetry points. (b) QSL bound state electron spectral
function $D_{\mathrm{QSL}}(\omega)$. Both (a) and (b) are under local
interaction strength $V_{r}=0.225\ \mathrm{eV}$.}

\label{QSL bound SD and SF} 
\end{figure}

\noindent From this, the gauge field corrected electron spectral density
can be obtained as $A_{\mathrm{QSL}}(\omega,\boldsymbol{k}_{1})=-\frac{1}{\pi}\mathrm{Im}G(c,c^{\dagger},\omega+i0^{+},\boldsymbol{k}_{1},\sigma)$,
and its spectral function $D_{\mathrm{QSL}}(\omega)=\frac{1}{N}\sum_{\boldsymbol{k}_{1}}A_{\mathrm{QSL}}(\omega,\boldsymbol{k}_{1})$,
as shown in Fig.~\ref{QSL bound SD and SF}.

In terms of QSL with Anderson lattice, by substituting the complete
spinon and chargon Green's functions into Eq. \eqref{P BSE}, and
define $G_{2}\left(P_{1}\right)=-G(\psi_{e},\psi_{e}^{\dagger},i\omega_{n},\boldsymbol{k},\sigma)$,
we can obtain the Eq. \eqref{electron GF} discussed earlier.

Those full Green's functions in Eq. \eqref{P BSE} are:

\noindent 
\begin{equation}
G^{-1}(\psi_{s},\psi_{s}^{\dagger},i\omega_{n},\boldsymbol{k},\sigma)=\left(\begin{array}{cc}
G_{0}^{-1}(a,a^{\dagger},i\omega_{n},\sigma) & -w\\
-w & G_{0}^{-1}(f,f^{\dagger},i\omega_{n},\boldsymbol{k},\sigma)
\end{array}\right),
\end{equation}

\noindent 
\begin{equation}
G^{-1}(Z_{c},Z_{c}^{\dagger},i\nu_{n},k)=\left(\begin{array}{cc}
G_{0}^{-1}(Y,Y^{\dagger},i\nu_{n}) & u\\
u & G_{0}^{-1}(X,X^{\dagger},i\nu_{n},\boldsymbol{k})
\end{array}\right).
\end{equation}

\begin{figure}
\noindent \begin{centering}
\includegraphics[width=0.8\linewidth]{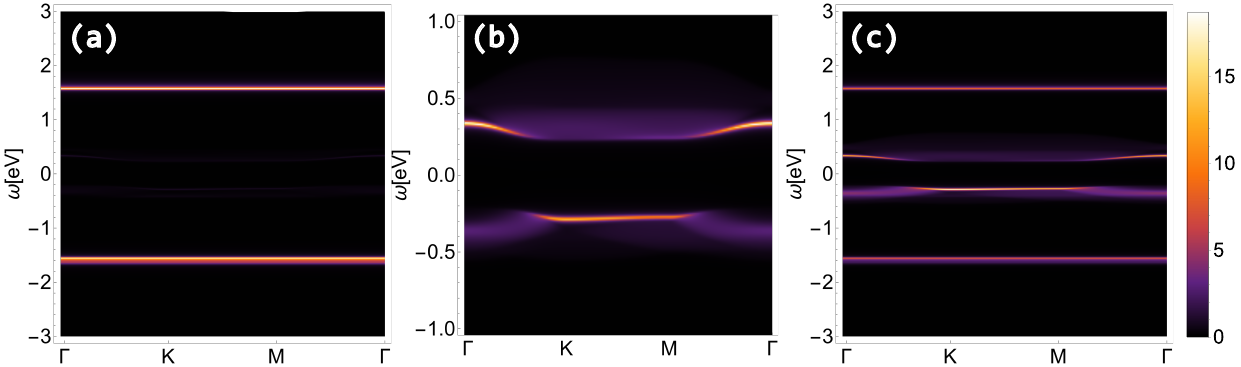} 
\par\end{centering}
\caption{(a), (b), (c) are spectral density of electron in spinon Kondo lattice
phase along the high symmetry points for Anderson lattice part, host
spin liquid part and the whole system together respectively, all with
the interaction strength $V_{\bm{r}}=0.225\ \mathrm{eV}$.}

\label{Electron spectral density} 
\end{figure}

\noindent Similar to Eq. \eqref{c BSE GF}, the expression for the
electron Green's function in the QSLAL can be readily obtained:

\begin{align}
G(\psi_{e},\psi_{e}^{\dagger},i\omega_{n},\boldsymbol{k}_{1},\sigma)= & -\frac{1}{\beta N}\sum_{i\nu_{n},\boldsymbol{k}_{2}}G(\psi_{s},\psi_{s}^{\dagger},i\omega_{n}+i\nu_{n},\boldsymbol{k}_{1}+\boldsymbol{k}_{2},\sigma)\otimes G(Z_{c},Z_{c}^{\dagger},i\nu_{n},\boldsymbol{k}_{1})\nonumber \\
 & -\frac{1}{\beta N}\sum_{i\nu_{n},\boldsymbol{k}_{2}}G(\psi_{s},\psi_{s}^{\dagger},i\omega_{n}+i\nu_{n},\boldsymbol{k}_{1}+\boldsymbol{k}_{2},\sigma)\otimes G(Z_{c},Z_{c}^{\dagger},i\nu_{n},\boldsymbol{k}_{1})\nonumber \\
 & \times i\nu_{n}\frac{V_{\bm{r}}}{U_{QSL}}G(\psi_{e},\psi_{e}^{\dagger},i\omega_{n},\boldsymbol{k}_{1},\sigma).
\end{align}
As an analogy of Eq. \eqref{c BSE GF} The simplified corrected electron
Green's function for QSLAL is given in Eq. \eqref{electron GF}

\noindent and their spectral function $A_{\mathrm{AL/QSL}}(\omega,\boldsymbol{k}_{1})$
are shown in Fig.~\ref{Electron spectral density}.

\end{document}